\documentclass[12pt,preprint]{aastex}
\catcode`\"=\active\let"=\"

\def\3{{\ss} }

\def\c12{{1\over 2}}

\def\plusplus{\raise 0.3ex\hbox{${\scriptstyle ++}$}{}}

\newcommand{\oversim}[2]{\protect{\mbox{\lower0.5ex\vbox{%
   \baselineskip=0pt\lineskip=0.2ex
   \ialign{$\mathsurround=0pt #1\hfil##\hfil$\crcr#2\crcr\sim\crcr}}}}} 
\newcommand{\simgreat}{\mbox{$\,\mathrel{\mathpalette\oversim>}\,$}} 
\newcommand{\simless} {\mbox{$\,\mathrel{\mathpalette\oversim<}\,$}} 

\begin{document}

\title{A comprehensive model for the Monoceros tidal stream}
\author{J. Pe\~narrubia$^1$, D. Mart\'inez-Delgado$^1$, H.W. Rix$^1$, M.A G\'omez-Flechoso$^2$, J. Munn$^3$, H. Newberg$^4$, E.F. Bell$^1$, B. Yanny$^5$, D. Zucker$^1$, E. K. Grebel$^6$}
\affil{$^1$Max Planck Institut f\"ur Astronomie, K\"onigstuhl 17, Heidelberg, D-69117, Germany}
\affil{$^2$ Universidad Europea de Madrid, Villaviciosa de Od\'on, E-28670, Spain }
\affil{$^3$ US Naval Observatory, Flagstaff Station, P.O. Box 1149, Flagstaff, AZ 86002, USA}
\affil{$^4$ Dept. of Physics \& Astronomy, Rensselaer Polytechnical Institute, Troy, NY 12180,USA}
\affil{$^5$ Fermi National Accelerator Laboratory, Batavia, IL 60510, USA}
\affil{$^6$ University of Basel, Venusstrasse 7, Binningen, CH-4102, Switzerland }

\email{jorpega@mpia-hd.mpg.de}

\begin{abstract}
We have compiled an extensive dataset on potential parts of the Monoceros tidal 
stream and performed an exhaustive survey of dwarf galaxy semi-analytic orbits in order
to constrain its orbital properties. The best-fit orbits are subsequently
realized as self-consistent $N$-body simulations in order to reproduce the
spatial and velocity distribution of satellite debris. We find that all
kinematic and geometric constraints can be fit by a single stream allowing
for multiple wraps. The orbital eccentricity and inclination of the
progenitor are strongly constrained to be $e = 0.10 \pm 0.05$ and $i = 25^\circ
\pm 5^\circ$. Ten new estimates of proper motions from the Sloan Digital Sky
Survey (SDSS) clearly exclude all retrograde orbits. 
Particles lost by the
satellite populate two, nearly concentric rings naturally explaining the
detection of stream stars at both 6--8\,kpc (Ibata et al. 2003, Newberg et al. 2003) and 
12--18 kpc (the Tri/And stream; Rocha-Pinto et al. 2003) from the Sun. 
We have attempted to predict the present location of the Monoceros stream progenitor using different information: (i) the kinematical and spatial distribution of detections and (ii) the different mean metallicity in the innner and the outer rings. Due to the lack of observational data in the whole range of Galactic latitudes, the geometrical/kinematical constraints leads to a wide range of possible locations. By associating older parts of the model stream with lower metallicity parts of the observed data, we argue in
favor of a current location of $l \sim 245, b \sim -18$., with a distance to the Sun $r_s\simeq $15 kpc. The mass of the progenitor has been poorly constrained due to the slow orbital decay. Similar fits have been obtained for masses $(3-9)\times 10^{8} M_\odot$. \\
We have analyzed to possible common origin of the Canis Major dwarf and the Monoceros stream. Canis Major dwarf moves on a prograde, nearly circular orbit ($e\simeq 0.16$) in the Milky Way disk ($i\simeq 4^{+14}_{-4}$ deg.). This orbital inclination is too low to account for the large vertical dispersion of stream stars. However, the bimodal distribution of radial velocities in the central region found by Martin et al. (2004b) probably indicates that their selection criteria to indentifying dwarf stars lead to a contamination of background stars. In that case, the kinematical data outlined above might result to an underestimated orbital inclination. Lastly, the distance estimation to Canis Major dwarf is around a factor two smaller than that obtained from our model. Unfortunately, the possible identification of the Monoceros stream progenitor in Canis Major remains unclear. 

\end{abstract}
\keywords{galaxies: evolution --- galaxies: formation ---galaxies: haloes ---
 galaxies: structure}

\section{Introduction}
In a $\Lambda$CDM universe, the inner regions of massive galaxies like the Milky Way gain a large fraction of their mass through tidal disruption and accretion of a large number of low-mass
fragments (e.g van den Bosch, Tormen \& Giocoli 2004). The fossil records of these merging processes may be observable nowadays in form of long
tidal streams or large-scale stellar substructures around the parent spiral galaxies. Numerical galaxy formation simulations show that dynamical friction brings sub-structures from outer halo regions to the neighbourhood of the parent galaxies' disks. If haloes are flattened (oblate) and their axi-symmetry plane is that of the disk, orbits of non-polar satellite galaxies tend to become co-planar and circularise with time (Pe\~narrubia, Kroupa \& Just 2004). In this scenario, tidal debris of several disrupted satellite galaxies might have contributed to the formation of the stellar disk if they followed nearly circular orbits with a low orbital inclination at late times of their evolution (Navarro 2004 and references therein). 

The Milky Way is an important laboratory in which the predictions of this cosmological scenario can be tested. In the last decade, large-scale surveys have proved the existence of tidal streams  (Sgr: Ibata et al. 1994; Monoceros: Newberg et al. 2002) in our Galaxy, providing strong observational evidence that disruption of dwarf satellites contributes to the assembly of some components of our Galaxy. These tidal streams offer an unique opportunity to study accretion events in considerable detail using the chemical, kinematic and spatial distribution of tidal stream's stars, which can be directly compared against predictions of
N-body simulations of merging events (e.g Law, Johnston \& Majewski 2004).

Recently, the Sloan Digitized Sky Survey team reported the discovery of a coherent ring-like structure  at low galactic latitude spanning about 100 degrees in the sky (Newberg et al. 2002; Yanny et al. 2003). Follow-up  observations (Ibata et al. 2003) found that this structure of low metallicity stars surrounds the Galactic disk at Galactocentric distances from $\sim$ 15 kpc to $\sim 20$ kpc. Different groups proposed two different scenarios to explain the nature of this feature in the outer Galactic disk (see Helmi et al. 2003): i) a new tidal stream (Yanny et al. 2003); and ii) a stellar ring resulting from perturbations in the disk similar to the remnants of ancient warps (Ibata et al. 2003). 
Tracing this structure with 2MASS M giant stars,  Rocha-Pinto et al. (2003) concluded that its structural characteristics are consistent with the interpretation of this structure as the fossil of a merging dwarf galaxy, similar to the Sgr dwarf galaxy (Ibata et al. 1994), in the Galactic plane.  Interestingly, Frinchaboy et al. (2004) and Crane et al. (2003) also suggested some nearby Galactic open and globular clusters with coordinated heliocentric radial velocities, indicating a possible common origin with the tidal stream.

 Unlike the Sagittarius dwarf galaxy, the Monoceros stream has been detected prior to locating the main body of the parent galaxy. The location is still controversial. From the analysis of the 2MASS catalogue, Martin et al. (2004a) reported a strong elliptical-shaped stellar over-density in the constellation of  Canis Major, which is coincident in position and radial velocity with a small group of 4 globular clusters (see Sec.~\ref{sec:cluster}). Using a fairly simple model, they identify the CMa over-density as the main body of the progenitor dwarf galaxy of the Monoceros stream (named Canis Major dwarf). Bellazzini et al. (2004) presented colour-magnitude diagrams in the
surroundings of the CMa location, concluding that the system is
situated at 8$\pm$ 1 kpc from the Sun and it is composed by a metal-rich, intermediate-age population. 
Momany et al. (2004), however, comparing proper motions and radial velocities of Canis Major M-giant stars, obtain that the overdensity in this region mimics the thick disk kinematics. Moreover, they find that the star counts in that region are those expected in the standard Milky Way model if taking into account the warp and flare of the disk.
However, a deep colour-magnitude diagram of the center of the Canis Majoris
over-density by Mart\'\i nez-Delgado et al. (2004) shows a very well-defined
main-sequence consistent with a limited extent in distance, confirming that
this stellar population is associated with a distinct, possibly still bound stellar system with properties (surface brightness, absolute magnitude,
stellar content) compatible with those of Local group dwarf spheroidal galaxies.
In base of these last results, the Canis Major overdensity will be referred to the CMa dwarf galaxy in this work.

In this paper we present the results of an extensive search of possible disruption scenarios of satellite galaxies to explore: a) how many of the 'over-density signatures' can be attributed to a single stream and b) to constrain its progenitor orbit. With an objective criterion we select those orbits that reproduce the spatial and kinematical distribution of candidate debris. Beyond exploring the possible progenitor location, the results will: (i) help to carry out new surveys in different regions of the sky and (ii) constrain the thick disk formation history through the distribution evolution of stripped stars.

This contribution has been organised as follows: In Section~\ref{sec:obs} we compile the observational data available so far. Section~\ref{sec:tools} outlines the galaxy and satellite models that we use in our investigation. We also describe our semi-analytic orbit algorithm and the results of the orbital parameter survey. We obtain the distribution of debris via the N-body algorithm sketched in Section~\ref{sec:nbody}. In Section~\ref{sec:res} we compare the best-fit orbit with the observational data, whereas in Section~\ref{sec:dis} we comment different implications of our results.

\section{Observational data}\label{sec:obs}
For the present work, we have attempted a comprehensive compilation
  of observational constraints on the geometry and kinematics of
  the Monoceros stream. Given the great deal of activity in detecting
  tidal streams in general, given the diversity of available data sets, and
  given in particular the great interest in the Monoceros stream spawned by
the
  original discovery, the constraints are both numerous and inhomogeneous.

  Typically, detections of overdensities have resulted from photometry,
  finding an axcess of main-sequence stars or M-giants
  at a given apparent magnitude. Depending on the tracer population, this
  approach provides distance estimates of differing quality.
  Radial velocities are available for a sub-set of stars within a sub-set
  of directions with detected overdensities. 
In principle, also the
  chemical composition of stars can serve as a constraint, as it can be
  expected to vary continuously along the stream, presumably with
  decreasing metallicity in the most loosely bound (and hence first
  lost) material.
Finally, proper motions
  (which typically can be measured to $\Delta\mu\sim$ 3-4 mas/year) can provide some
  rough contraints; their precision in physical units is, however, only
  $\Delta v_{\rm perp} = \Delta \mu\times r_s \times 4.74$, where $r_s$ is the distance from the sun and 4.74 is the factor that coverts [kpc mas/yr] into [km/s]. 

  We have compiled both published constraints and those available to us,
  but still in the process of publication. A few constraints were derived
  specifically for the present paper. We have summarized the constraints,
  their nature (photometric,kinematic, etc..) and their sources in Table~\ref{tab:obser}. In following plots we use different symbols in order to distinguish between different data sources. In order to simplify our Figures we have made an exception for those data with available radial velocities, which we have been plotted everywhere with {\em full triangles}. 
\footnotesize
\begin{table}
\begin{center}
\caption{Observational constraints}
\begin{tabular}{||l |l  |l |l |l |l |l   ||} 
\tableline \tableline
Authors  &$l-b$ range & Type  & $v_r^{[1]}$ & $\mu^{[2]}$ & $N^{[3]}$ & Symbol \\ \tableline
Yanny et al. (2002);    &$[182^\circ,225^\circ]$, $[+28^\circ,-27^\circ]$ & CMD          & yes & no & 4 & {full triangle} \\
Ibata et al. (2003)       &$[122^\circ,218^\circ]$, $[+30^\circ,-25^\circ]$ & CMD$^{[4]}$       & no  & no & 14 &{open triangle} \\
Crane et al. (2003)       &$[157^\circ,242^\circ]$, $[+38^\circ,-15^\circ]$ & M-giant       & yes & yes$^{[6]}$  & 58  &{full triangle} \\
Newberg et al. (2004)     &$[110^\circ,225^\circ]$, $[+37^\circ,-32^\circ]$ & CMD$^{[4]}$       & no  & no & 22 &{open box} \\
Rocha-Pinto et al. (2004) &$[117^\circ,157^\circ]$, $[+38^\circ,-25^\circ]$ & M-giant$^{[5]}$   & yes & no & 31 &{full box} \\
\tableline
\end{tabular}
\label{tab:obser}
\tablenotetext{}{$^{[1]}$ Radial velocity measurements. \\ $^{[2]}$ Proper motion measurements. \\ $^{[3]}$ Number of detections\\ $^{[4]}$ Possible confusion with thick disk stars (see Section~\ref{sec:disk}). \\ $^{[5]}$ Tri/And stream.\\$^{[6]}$ Only for ten stars.}
\end{center}
\end{table}
\normalsize

 Additionaly, we have proper motions of confirmed stream star members selected from
the radial velocity compilation by Crane et al. (2003).
Proper motions are determined by combining recalibrated USNO-B1.0
 positions (Monet et al. 2003) with those from the Sloan Digital Sky Survey \footnote{SDSS (York et al. 2000) is an imaging and spectroscopic
survey that will eventually cover $\sim 1/4$
of the sky.
Drift-scan imaging in the five SDSS bandpasses ($u,g,r,i,z$)
(Fukugita et al. 1996, Gunn et al. 1998, Hogg et al. 2001) is processed through data reduction pipelines
to measure photometric and astrometric properties
(Stoughton et al. 2002, Smith et al. 2002, Pier et al. 2003, Abazajian et al. 2004) and
to identify targets for spectroscopic followup.} catalog, as
detailed in Munn et al. (2004).
Proper motion of our target stars are given in Table~\ref{tab:obs}.
It is necessary to clarify that the measurements of proper motions have not
been included in the orbital fit and will be used only to discriminate the sense
of motion of the progenitor's orbit. 

\begin{table}
\begin{center}
 \caption{Proper motions of Monceros stream stars}
\begin{tabular}{||l |l  |l |l |l |l   ||} 
\tableline \tableline
{\bf l}  &{\bf b}  & {\bf $r_s$} & {\bf $v_r$} & {\bf $\mu_l$} & {\bf $\mu_b$}  \\
(deg.)   & (deg.)  &  (kpc)      &  (km/s)     & (mas/yr)& (mas/yr) \\ \tableline
186.382 &  23.910 &  11.2 &  $-6.8\pm  3.9$ & 2.6 & -1.9 \\
189.316 &  23.251 &  12.6 &  $149.4\pm 5.6$ & 0.6 & -4.4 \\
186.894 &  24.181 &  10.0 &  $5.4  \pm 3.7$ &-5.5 & -13.9 \\
189.741 &  23.345 &  12.7 &  $0.2  \pm 6.0$ & 6.0 & -1.0 \\
198.778 &  25.063 &  11.9 &  $49.1 \pm 2.5$ &-1.2 &  2.6 \\
224.108 &  21.344 &  10.9 &  $82.8 \pm 3.4$ & 1.0 & -6.3 \\
178.371 &  36.786 &  10.4 &  $-19.7\pm 2.0$ & 6.1 &  1.5 \\
186.992 &  38.816 &  12.3 &  $42.3 \pm 2.1$ & 0.4 &  3.0 \\
221.989 &  29.900 &  12.3 &  $86.1 \pm 3.2$ &-3.3 & -6.2 \\
223.126 &  32.262 &  11.8 &  $55.1 \pm 2.7$ &-3.5 &  1.9 \\
\tableline
\end{tabular}
\label{tab:obs}
\tablenotetext{}{Positions in Galactocentric coordinates of the Monoceros stream stars with measured proper motions. The errors in the heliocentric distance $r_s$ are about 25\% of the value, whereas in $\mu_l,\mu_b$ they have been estimated to be 3.5 mas/yr.}
\end{center}
\end{table}

  In addition, we are faced with the problem of which constraints (on which
  candidate sections of the Monoceros stream) to include in our modelling.
  Guided by the goal of finding the largest number of stream portions, that may
have
  arisen from a single disruption event, we have used an iterative
  procedure.
  Starting with the original stream detections (Newberg et al. 2002.,Yanny  et al. 2003)
  and the comprehensive kinematic survey of M-stars (Crane et al. 2003) spanning 100$^\circ$ in the sky.
Initial modelling of these constraints made it clear, that other,
independently
  found, overdensities, most likely are also part of the stream. In particular, the Tri/And stream (Majewski et al. 2004), a more distant metal-poor stellar stream, showed a location in the sky and a radial velocity curve fairly similar to the predictions of our first-iteration model. Therefore, the available data (including radial velocities from Rocha-Pinto et al. 2004) on the Tri/And stream were then included as inputs in a second iteration of our survey of the best-candidate orbit to better constrain its properties and reduce the number of possible scenarios.

  Finally, we explicitly test whether the Canis Majority dwarf
  is likely part of of the stream, presumably the 'parent' of the tidal debris.
  We do this by omitting CMa in the dirst modelling, and then comparing its orbit with that of our model for a possible Monoceros stream progenitor.

\section{Fitting numerical models to observational data}\label{sec:tools}
In this Section we outline how we perform the orbital analysis of dwarf satellites and the method used to fit numerical orbits to the observational data. \\
We first describe a Milky Way model that matches the mass distribution of our Galaxy. Subsequently, we discuss the parameter space that we must cover in order to determine the orbital characteristics of a possible Monoceros stream progenitor.
Lastly, we present the method used to constrain the orbital properties of a possible Monoceros tidal stream progenitor, which can be divided in two steps: (i) Satellite orbits are calculated using a well-tested semi-analytic algorithm in order to perform a survey of our large parameter space. Subsequently, the orbit collection has been compared against the observational sample to determine the parameter sets that provide the best fits.  (ii) Once the orbital constraints are known, we carry out N-body simulations to analyze the spatial and kinematical distribution of debris.  

\subsection{Galaxy and dwarf satellite models}\label{sec:models}
Our dynamical model for the Milky Way follows Hernquist (1993), where: (i) the disk is exponential in the radial direction and isothermal in the vertical direction; (ii) the bulge is described by a Hernquist model (Hernquist 1990) with spherical symmetry; (iii) the halo follows a flattened, non-singular isothermal profile with given core and cut-off radius.\\

For the density distributions of the disk we take
\begin{equation} 
        \rho_d(R,z)=\frac{M_d}{4\pi R_d^2 z_0} \rm{exp}(-R/R_d) 
        \rm {sech}^2(z/z_0),
\end{equation}
where $M_d=5.60\times 10^{10} M_\odot$ is the disk mass, $z_0=0.70$ kpc is the vertical thickness, and $R_d=3.50$ kpc is the
exponential scale length in the radial direction. The mass profile
decays exponentially with R and is composed of isothermal sheets along
the vertical direction.  Velocities are assumed to have a Gaussian
distribution.  

For the bulge we adopt the spherical Hernquist profile (Hernquist 1990),
\begin{equation} 
        \rho_b=\frac{M_b}{2\pi} \frac{a}{r(r+a)^3},
\end{equation}  
where $M_b=1.86\times 10^{10} M_\odot$ is the bulge mass and $a=0.53$ kpc is the spherical scale length. This
analytical profile fits the de Vaucouleurs law (de Vaucouleurs 1948).
The velocity field is constructed from the Jeans equations by assuming
isotropic Gaussian velocity distributions at each radial distance
(Hernquist 1993).

We use a non-singular isothermal profile for the dark matter halo (DMH),
\begin{equation} 
        \rho_h=\frac{M_h \alpha}{2\pi^{3/2}
        r_{\rm cut}}\frac{{\rm exp}(-m^2/r_{\rm cut}^2)}{m^2+\gamma^2},
\label{eqn:rho_h}
\end{equation}
where
\begin{equation} 
m^2\equiv R^2+z^2/q_h^2
\label{eqn:m}
\end{equation}
in cylindrical coordinates. $q_h$ is the halo density flattening, $M_h=7.84\times 10^{11} M_\odot$ the DMH mass, $r_{\rm cut}=84.00$ kpc the cut-off radius, $\gamma=3.50$ kpc
the core radius, and
\begin{eqnarray}
        \alpha\equiv\{1-\sqrt{\pi}\beta{\rm exp}(\beta^2)[1-{\rm erf}
                      (\beta)]\}^{-1} =  \\ \nonumber
                  1 + \sqrt{\pi}\beta + (\pi -2) \beta^2 + O(\beta^3) 
\end{eqnarray} 
where $\beta=\gamma/r_{\rm cut}\simless 1/24$  {in our
calculations}. To construct the flattened
(oblate) DMHs, a non-homologous transformation is applied to
(\ref{eqn:rho_h}) to achieve the desired axis ratio $q_h$ while 
preserving the central density. In order to minimise computational time when constructing flattened
DMHs with embedded bulges and disks, we apply a highly efficient
technique using multipole potential expansions to tailor the local
velocity ellipsoid to the required morphology (Boily, Kroupa \& Pe\~narrubia 2001). The algorithm to add together individual
components in a single galaxy is adapted from 
Hernquist (1993).  This code scales linearly with
particle number and hence we can construct flattened DMHs with $\simgreat10^6$ particles in a short computational time.   

The specific parameters are chosen to reproduce the observed rotational curve of the Milky Way. The only free parameter of the Galaxy model that we explore in our calculations is the halo axis-ratio ($q_h$).

The dwarf satellite model follows a King profile (King 1966) with central potential $\Psi/\sigma^2=4$ and concentration $c=\log_{10}(r_t/r_K)=0.84$, where $r_t,r_K$ are the tidal and King radii, respectively. 
Since we are assuming that the tidal debris represent multiple wraps (and thus, multiple perigalacticon passages) of the tidal stream originating from one disrupting satellite, we restrict the tidal radius of the progenitor dwarf galaxy in each model with the following criteria: (i) the tidal radius must be small enough to prevent galaxy disruption by the first peri-galacticon passage and (ii) the tidal radius must be large enough to induce a progressive mass loss that leads to the formation of a tidal stream.
We have selected the satellite tidal radius to match the Jacobi limit (Section 7.3 of Binney \& Tremaine 1986) at the first peri-galacticon, which is a simple method to achieve 'slow' mass loss along the orbit.

\subsection{Orbital parameters}\label{sec:paramsp}
Given the Milky Way and satellite mass profiles, three aspects determine the subsequent orbit evolution: (i) the flattening of the Milky Way halo $q_h$; (ii) the mass of the satellite, determining the degree of dynamical friction and (iii) the initial orbital geometry. The corresponding free parameters are outlined in Table~\ref{tab:param}. 
 We have performed 151,200 simulations by means of our semi-analytic code (see below) in order to fit the observational data. Each orbit is repeated with 12 different azimuthal angles. This turns out to be necessary since dynamical friction prevents an orbit from filling up its phase space surface which, therefore, ``breaks'' the axi-symmetry of the problem.

Due to the high efficiency of the semi-analytic code, the resolution achieved in the parameter space is constrained by the limted number of available observational data rather than by CPU limitations. More densely-sampled parameter surveys would not provide stronger constraints, owing to the large degeneracy seen already in this sparse sampling of parameters space (see Section~\ref{sec:res_sa}).

\begin{table*}
\begin{center}
\caption{Paramenter space}\label{tab:param}
\begin{tabular}{||l |l  |l |l |l |l |l  |l ||} \tableline \tableline
{\bf Parameters} & $M_s (\times 10^8~M_\odot)$                     & $r_a$ (kpc)&$e$        & $i$ (deg.)         & $\phi$ (deg.) & $ q_h $ &\\ \tableline
Range            & $[0.6 ,12.0]$ &[17.5,80.5]  &[0.,0.7] &[5.0,45.0]          &[0.0,360.0]    & [0.5,1] &\\
                 &                         &             &                      &[135.0,175.0]       &               &         &\\
Precision        & $\pm 3.0$        & $\pm 3.5$         &$\pm 0.05$       & $\pm 5.0$                & $\pm 15.$           & $\pm 0.05$    &\\\tableline 
\# values         & 3                       & 10          &7         & 10                 & 12            & 6       &\\\tableline\tableline
{\bf Best fits}  &    $M_s (\times 10^8~M_\odot)$      &$r_a$ (kpc)  & $e$      &   $i$ (deg.)       &  $\phi$ (deg.)& $ q_h $ &$\chi_{\rm best}$ (kpc) \\\tableline
Prograde  &&&&&&& \\
{\em pro1}         & $6.3$            & 22.8        &0.1       &25.0                &300.0          & 0.6     & 4.4\\
{\em pro2}         & $6.3$            & 22.8        &0.1       &25.0                &150.0          & 0.8     & 4.6\\
{\em pro3}         & $6.3$            & 22.8        &0.1       &25.0                &120.0          & 0.7     & 4.7\\\tableline
Retrograde  &&&&&&&\\
 {\em ret1}        & $12.0$            & 22.8        &0.5       &165.0               &180.0          & 0.6     & 5.6\\
 {\em ret2}        & $12.0$            & 22.8        &0.5       &155.0               &210.0          & 0.6     & 5.7\\
 {\em ret3}        & $12.0$            & 22.8        &0.5       &165.0               &210.0          & 0.5     & 5.7\\
\tableline 
\end{tabular}
\tablenotetext{}{All quantities are given at $t=0$ of the simulation. $r_a$ denotes the initial distance to the Galaxy centre, $e\equiv (r_a-r_p)/(r_a+r_p)$ the orbital eccentricity, $r_a, r_p$ being the apo and peri-center distances, $i$ is the inclination with respect to the disk plane, $\phi$ is the azimuthal angle in spherical coordinates and $q_h$ is the axis-ratio of the halo's density profile. Values of free parameters are equally distributed with the ranges. Our notation is so that $0^\circ<i<90^\circ$ indicates a prograde motion whereas $90^\circ<i<180^\circ$ a retrograde one.}
\end{center}
\end{table*}

\subsection{Semi-analytic fit}\label{sec:fit_sa}
Exploring a large number of initial conditions and accounting for the time-dependent dynamical friction and satellite disruption can only be carried out in a reasonable time by using semi-analytic algorithms. We have used that proposed by Pe\~narrubia (2003) which provides, for the galaxy density profile outlined in Section~\ref{sec:models}, the evolution of ${\bf r}$, where ${\bf r}$ is the satellite centre-of-mass position, and the dwarf satellite mass $M_s$. This code has been tested against N-body calculations for a large spectrum of orbital parameters and satellite masses (Just \& Pe\~narrubia 2004) as well as halo flattenings (Pe\~narrubia, Just \& Kroupa 2004), showing that $|{\bf r}_{\rm analytic}-{\bf r}_{\rm Nbody}|\leq 0.7$ kpc for time integrations of 3 Gyr, once the Coulomb logarithm is fit to N-body simulations (the best-fit values being $\ln \Lambda_h=2.1$ for the halo and $\ln \Lambda_d=0.5$ for the disk). 

 In order to compare our collection of orbits to the observational data we assume that the distribution of stream stars can be reproduced by a single stream within a given number of wraps. Equivalently, we can define the number of wraps as the time interval that a point-mass particle needs to cover the phase space defined by the observational data, $\Delta T$, which {\em a priori} is an unknown quantity that we have estimated from our N-body simulations to lie between $2T_{\rm orb}\leq \Delta T\leq 3 T_{\rm orb}$, where $T_{\rm orb}=2 \pi r/v_c$ is one dynamical period at $r\sim 20 $kpc from the Galaxy centre; $v_c=220 $km~s$^{-1}$. The maximum integration time of semi-analytic orbits is $T_{\rm max}=5T_{\rm orb}\simeq 3$ Gyr.

The fitting algorithm that we use is the following:\\
(i) Each orbit is divided into $n$ overlapping segments of extension $\Delta T$. Following the above discusion, we carry out the fit for stream extensions of $\Delta T=1.2, 1.8$ Gyr (i.e  $\Delta T\simeq 2,3 T_{\rm orb}$). The number of segments for a single orbit is $n=4,3$, respectively\footnote{This selection leads to fixed overlapping time intervals of extension $(n\Delta T-T_{\rm max})/(n-1)$, i.e, 0.6 Gyr for $n=4$ and 1.2 Gyr for $n=3$.}.\\
(ii) For a given segment we calculate $\chi_{i,j}^2=({\bf r}_{\rm j, analytic}-{\bf r}_{\rm i,obs})^2+ K (v_{\rm j, rad, analytic}-v_{\rm i, rad, obs})^2$, where $v_{\rm rad}$ is the heliocentric radial velocity and $K$ determines the relative weights of kinematical and spatial constraints ($K\simeq 0.2$ kpc km$^{-1}$ s). The sub-indexes $i,j$ take the values $i=1,2,..., N_{\rm obs}$, where $N_{\rm obs}$ is the number of observational data, $j=1,2,..., N_{\rm analytic}$, and $N_{\rm analytic}$ is the number of semi-analytic points in a segment. Subsequently, for each observational value $i$ we look for the semi-analytic point that leads to the smallest value of $\chi_{i,j}$, obtaining a set of $N_{\rm obs}$ semi-analytic points where the $\chi-$values find a minimum. The average value of $\chi$ in a given segment is $\chi=\sum \chi_{i}/N_{\rm obs}$.  \\
(iii) The calculation is repeated for the rest of segments. \\
(iv) The $\chi$-values of each segment are sorted, obtaining the time interval of the orbit where the best fit occurs and the minimum $\chi$. We must note that this method fits observational to semi-analytic points, and not {\em vice versa}, which allows us to fit more than one orbital period to observational points located in a small region of the space (see Section~\ref{sec:res_sense}).\\
(v) The process repeats for the whole orbit collection (regarding that each orbit corresponds to one point of our parameter space).

\subsection{N-body code}\label{sec:nbody}
The semi-analytic code reproduces remarkably well the motion of a dwarf satellite centre-of-mass and its mass evolution. However, as Piatek \& Pryor (1995) showed, the mass loss process itself is fairly complicated and difficult to implement in semi-analytic algorithms.

In order to describe the distribution of stripped stars in the Galaxy we perform N-body simulations from the best-fitting orbits found using the semi-analytic code.
We carry out self-consistent N-body calculations with {\sc superbox}, a particle-mesh code (see Felhauer et al. 2000), which calculates the gravitational potential in three boxes centred at the disk, bulge, halo and at the dwarf satellite, each box with $64^3$ grid-cells. We refer the reader to Pe\~narrubia (2003) and Pe\~narrubia, Kroupa \& Boily (2002) for a detailed description of the code parameters. Here we merely comment that, after fixing the time-step to 0.65 Myr, we obtain a conservation of total energy and total angular momentum of around 1\%.

The number of particles of each sub-system is: $N_d=9.0\times10^4$ (disk), $N_b=5.0\times 10^3$ (bulge), $N_h=1.2\times 10^6$ (dark matter halo) and $N_s=1.0\times 10^5$ (dwarf galaxy).

As in the case of semi-analytic calculations, we evolve N-body satellites approximately 3 Gyr. Three factors induce lead us to select this integration time: (i) to reproduce semi-analytic calculations thoroughly, (ii) to minimise feedback effects from the host galaxy affecting real dwarf satellites that are difficult to implement in N-body realizations of the Milky Way (such as the disk's spiral arms, over-density regions, warps ... etc). Those effects can be treated like small corrections to our Galaxy potential and, thus, neglected in short-time orbit calculations. And (iii) to approximate the orbits of tidal stream stars as the orbit of the main system. When obtaining the progenitor's main orbital properties from semi-analytic calculations, one assumes implicitly that the escaping particles follow the main system's orbit during a given time interval. This assumption only holds, therefore, for a limited number of orbital periods.  

\begin{figure}
\plotone{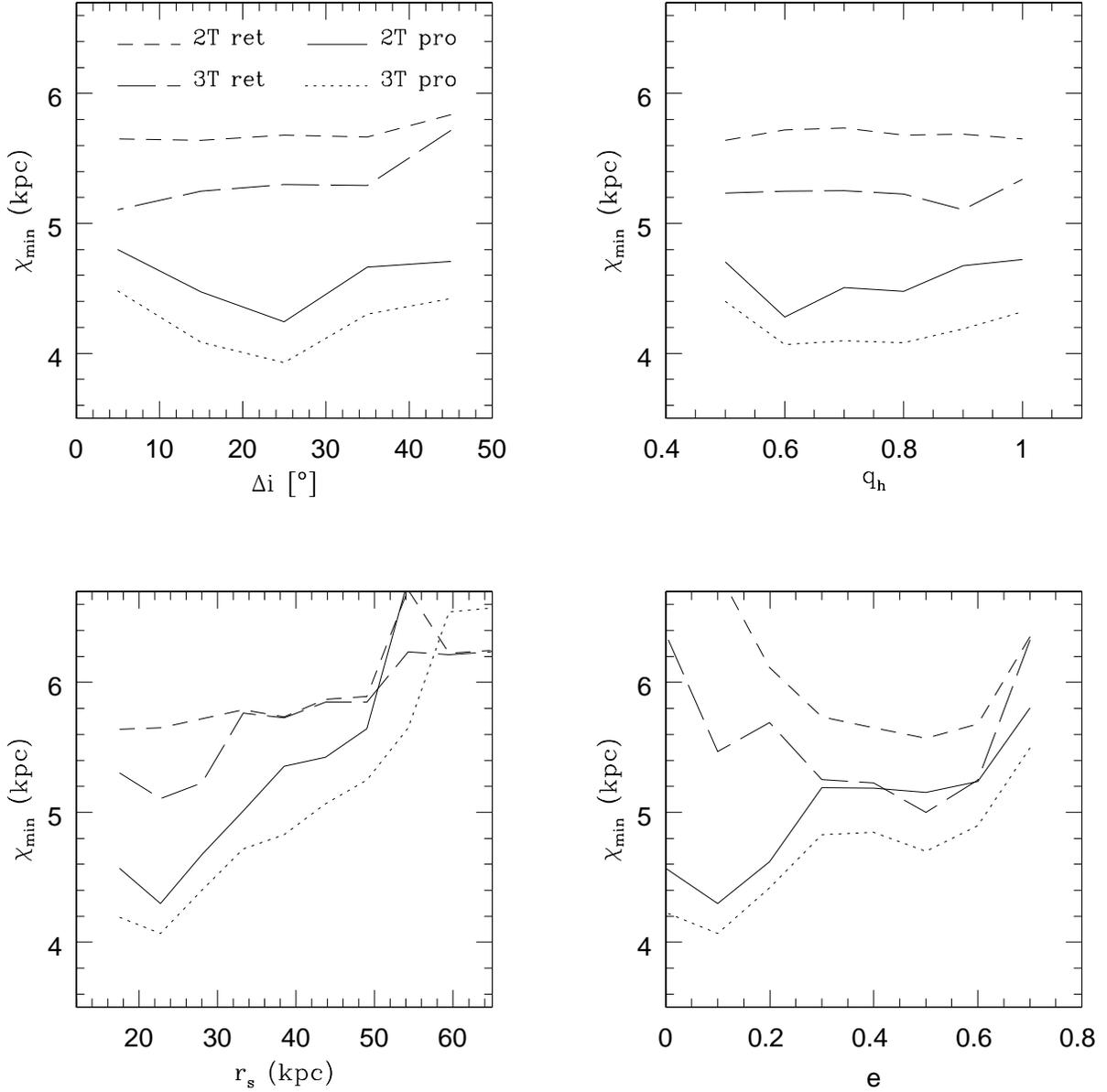}
\caption{ Minimum values of the fitting parameter $\chi$ as a function of initial heliocentric distance (lower-left panel), eccentricity (lower-right panel), halo axis-ratio (upper-right panel) and orbital inclination (upper-left panel). For each parameter, we perform the fit assuming that the observed tidal stream can be reproduced by an orbit segment of extension $\Delta T\simeq 2 T_{\rm orb}$ (solid lines) or $3 T_{\rm orb}$ (dotted lines).}
\label{fig:aj}
\end{figure}

\section{Results}\label{sec:res}

\subsection{Semi-analytic solutions}\label{sec:res_sa}

In Fig.~\ref{fig:aj} we plot the results of our semi-analytic fit to observational data for two different stream extensions, $\Delta T\simeq 2 T_{\rm orb}$ and $3 T_{\rm orb}$, and for prograde and retrograde orbits, as the sense of rotation has not yet been unambiguously determined\footnote{As Martin et al. (2004a) pointed out, the radial velocity curve as a function of projected position does not provide by itself sufficient information to determine whether the progenitor of the Monoceros stream follows a prograde or a retrograde orbit.}. 
This Figure shows that the best-fitting parameter values do not depend on the exact value of $\Delta T$. However, larger segments do alter the value of the minima and the smoothness of the curves by increasing the ratio $N/N_{\rm obs}$ (i.e, in a given segment, more semi-analytic points can be fit to the same number of observational constraints), improving the quality of the fit but, at the same time, leading to less pronounced minima, which hardens the selection of the best-fitting parameter sample.\\
 In lower-right panel we show the minimum value of $\chi$ as a function of the orbit's initial eccentricity. The minima are located at $e\simeq 0.1$ for prograde orbits and $e=0.5$ for retrograde ones. As we see in the lower-left panel, the initial heliocentric distance that leads to best fits independent of the orbital sense of motion is $r_s\simeq 22.8$ kpc. \\
In the upper-left panel we show the dependence of $\chi_{\rm min}$ on the inclination angle with respect to the disk plane (note that the orbital inclination of retrograde orbits is $i=180-\Delta i$ with this notation). We find that the orbital inclination of retrograde orbits cannot be accurately determined. These orbits reach the solar circle which, therefore, reduces the number of passages by the Galactic anticenter (where most of the observational data are located). As a result, retrograde solutions cannot account for the large vertical dispersion of observational points (see Fig.~\ref{fig:retsa}). In contrast, prograde orbits present a well-marked minimum at $i\simeq 25^\circ$. \\
The absence of a defined vertical structure of debris also leads to a degenerate value of the halo flattening, as we see in the upper-right panel. For prograde orbits, the halo axis-ratio that leads to the best solution is for $q_h=0.6$, although similar $\chi_{\rm min}$ have been found for $q_h=0.7,0.8$. No information about the halo flattening can be obtained from the fit if orbits are retrograde.

\begin{figure*}
\plotone{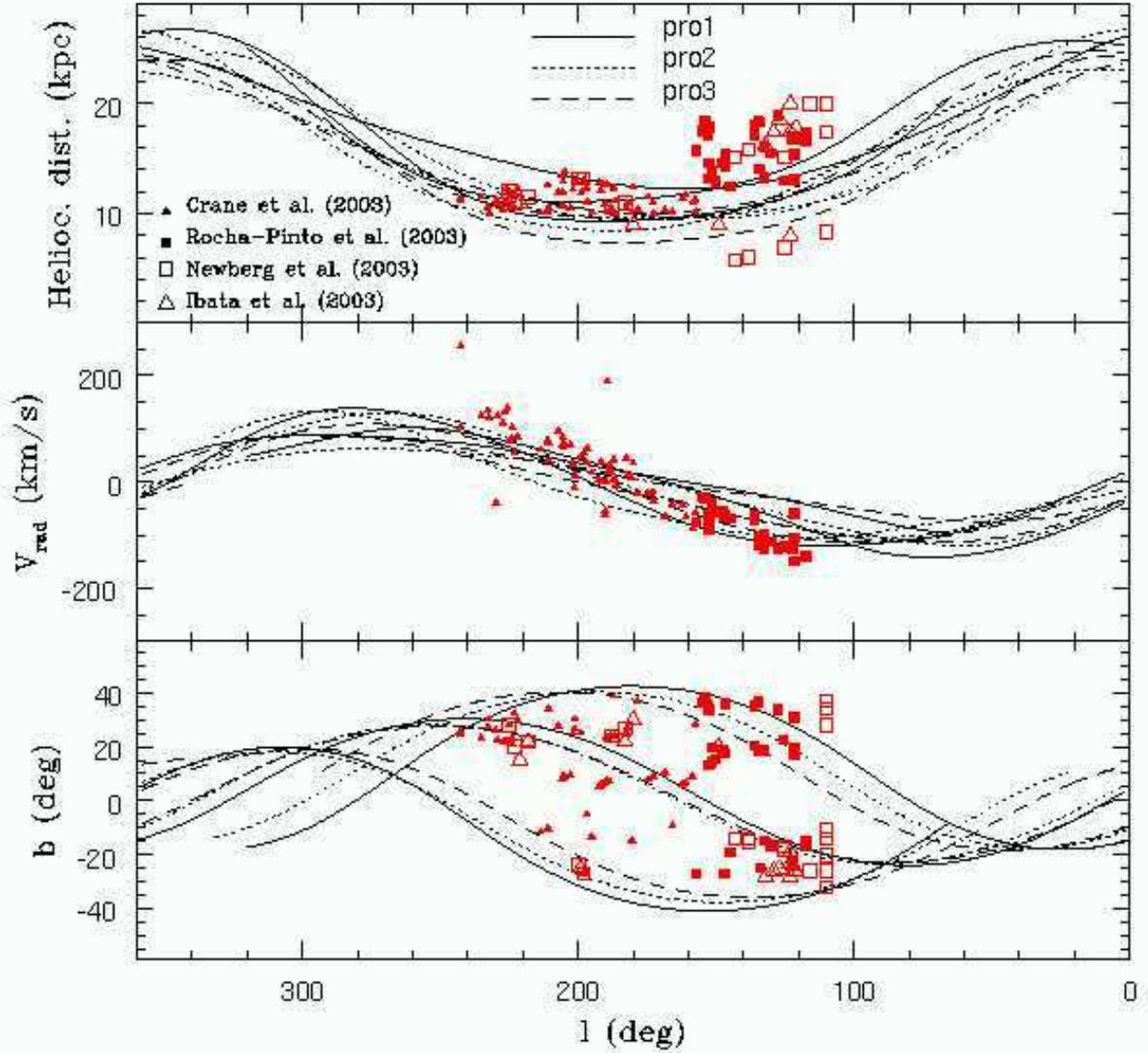}
\caption{ Best-fitting {\em prograde} solutions (see Table~\ref{tab:param}). Upper panel: Heliocentric distance as a function of Galactic longitude ($l$).  Middle panel: Heliocentric radial velocity. Lower panel: Projected spatial position in the $l-b$ plane. Open triangles and full squares represent detections of stars. Open squares denote colour-magnitude detections (with no radial velocity measurements available). }
\label{fig:prosa}
\end{figure*}

\begin{figure*}
\plotone{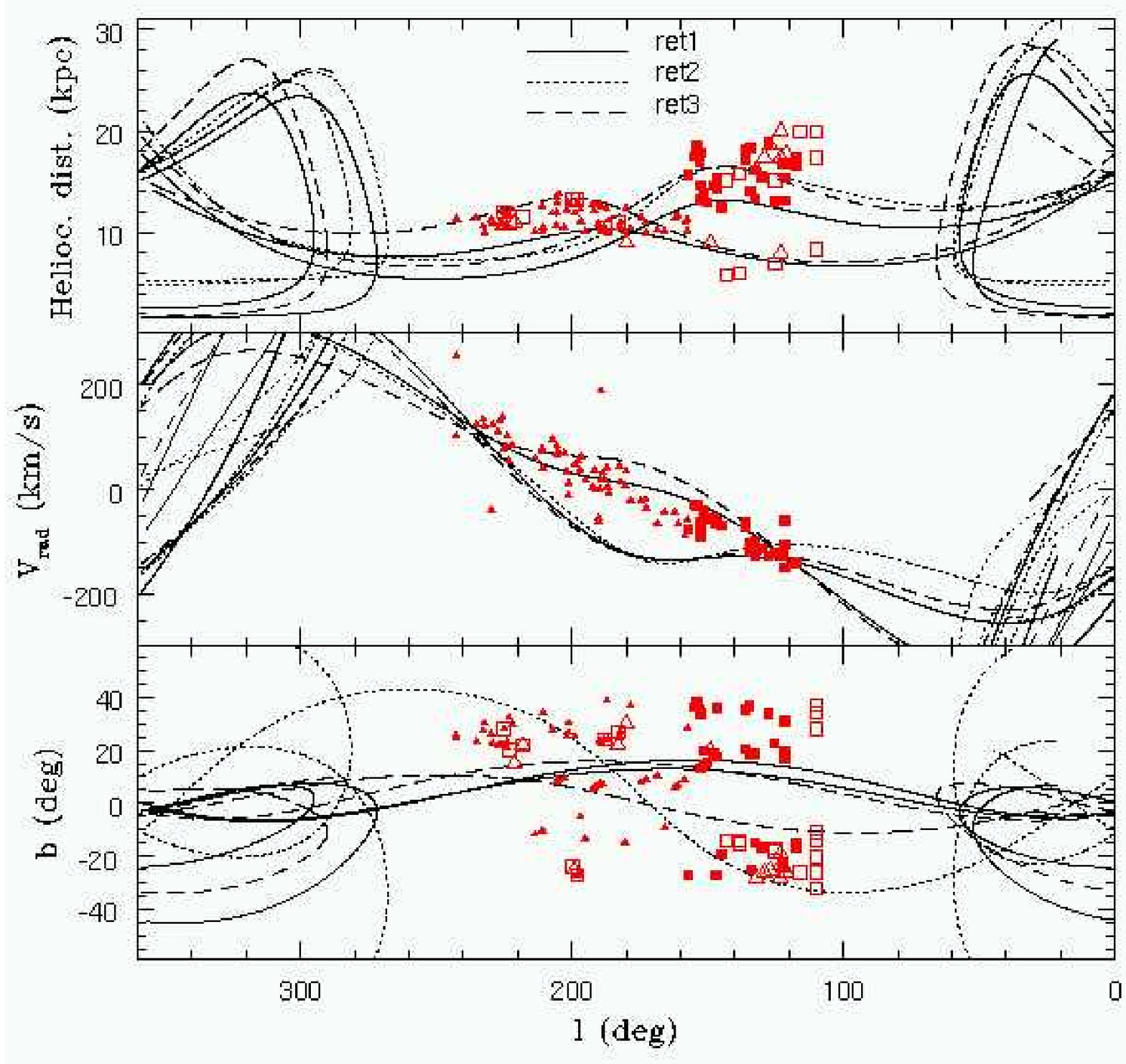}
\caption{ As Fig.~\ref{fig:prosa} for our best-fitting {\em retrograde} solutions.}
\label{fig:retsa}
\end{figure*}

As one would expect there are a large number of orbits that lead to similar values of $\chi$. That is far from surprising taking into account that: (i) the region of the Galaxy where the Monoceros stream has been detected is relatively small ($r_s\in [12,20]$ kpc, $110^\circ<l<240^\circ$, in galactocentric coordinates), which represents a small interval of the orbit and (ii) the stream presents a large dispersion in the $z$ direction (perpendicular to the disk plane) with a poorly defined structure (see Fig.~\ref{fig:prosa}).

In Table~\ref{tab:param} we summarize the fit results and show the three best-fitting retrograde and prograde orbits. The values of $\chi_{\rm min}$ clearly indicate that retrograde orbits lead to considerable worse  fits $\chi_{\rm ret}\simeq 1.3\chi_{\rm pro}$. 

 Our fit technique also provides a coarse estimate of the progenitor's initial mass, which we estimate to be around $6\times 10^8 M_\odot$ if it moves on a prograde orbit and $1\times 10^9 M_\odot$ if the orbit is retrograde. We remark that this result is fairly approximate (see precision estimates in Table~\ref{tab:param}), since the orbital decay within 3 Gyr is very low for prograde orbits and practically negligable for retrograde ones (note that the effect of the dwarf satellite mass on the semi-analytic orbit calculation occurs through dynamical friction). The poor sensitivity of the result to the satellite mass justifies our early choice of a coarse grid in satellite masses.

In Fig.~\ref{fig:prosa} we compare the semi-analytic satellite orbits with the observational data used for the fit. We only show the orbit at the time interval when the best fit to observations occurs (see Sec.~\ref{sec:fit_sa}).
The best prograde orbits stay relatively far from the Galaxy centre, $r \in [20,25]$ kpc, and has a peri-center at $l\simeq 180^\circ$. Different values of halo flattening and initial azimuthal angle match the observed heliocentric distances, radial velocities and projected positions of debris with a similar accuracy, leading to multiply denerate solutions. 
 
In Fig.~\ref{fig:retsa} we illustrate our best-fitting retrograde orbits. Those orbits have an apo-galacticon at $l\simeq 180^\circ$, where most of observational points are located. They reach the solar circle, moving within $r\in [7.5,22]$ kpc. Since orbits in the Galactic potential follow rosettes, the retrograde orbits pass the Galactic anticenter considerably less frequently than the prograde ones. 

 The satellite tidal radius was selected to match that of the Galaxy at the peri-centre, therefore, we find that for prograde satellites $(r_K,r_t)=(0.51, 1.79)$ kpc whereas for retrograde ones $(r_K,r_t)=(0.08, 0.54)$ kpc, where $r_K, r_t$ are the King and the tidal radius, respectively. Note that our retrograde satellites are approximately 3.3 times smaller than the progrades in order to prevent tidal disruption by the first peri-galacticon passages.

\subsection{N-body calculations. Orbital sense of motion}\label{sec:res_sense}
\begin{figure*}
\plotone{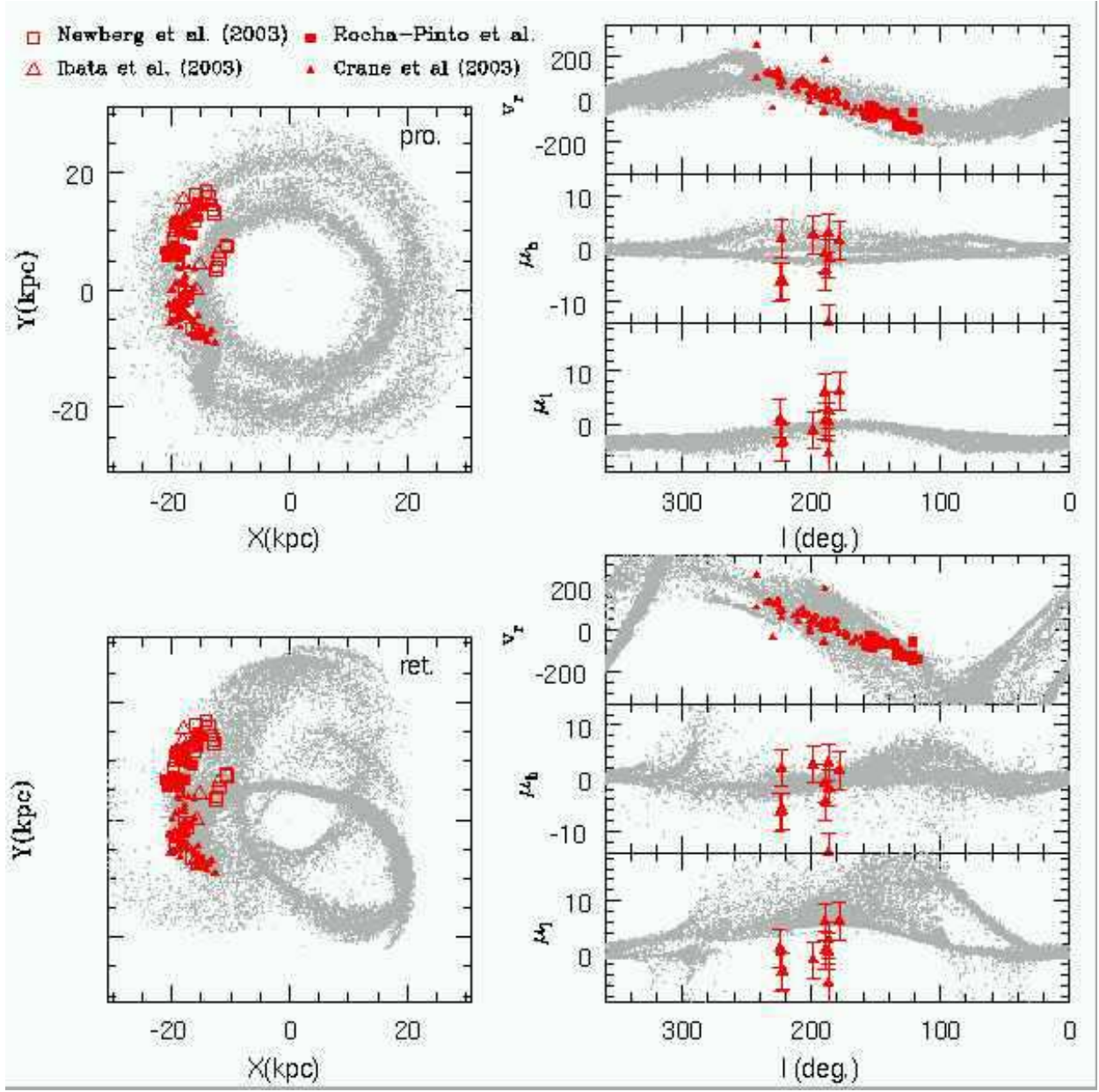}
\caption{ Prograde ({\em pro1} model) vs. retrograde ({\em ret1} model) orbits. Left column: X-Y Galaxy plane projections. The sun is placed at $(X,Y,Z)=(-8,0,0)$ kpc.  Right column: Heliocentric radial velocity curve (upper panel, given in km/s).  Proper motions in the latitudinal (middle panel) and in the longitudinal (lower panel) components, both given in mas/yr. }
\label{fig:proper}
\end{figure*}

In this Section we analyze the resulting kinematical and spatial distribution of debris from the best-fitting orbits obtained by our N-body algorithm.
In Fig.~\ref{fig:proper} we plot the projection of the pro and retrograde orbits in the X-Y plane, the radial velocity curve and proper motions (in Galactocentric longitudinal and latitudinal components) obtained from N-body realizations of our best-fit models and compare them to the observational data outlined in Section~\ref{sec:obs}. We have integrated the prograde orbit 2.99 Gyr in order to reproduce the observed projected distribution of debris (see Section~\ref{sec:mb}) whereas the retrograde one was evolved only 1.9 Gyr, until it was close to total disruption. Even imposing the initial satellite tidal radius to match the Jacobi limit at the first peri-galaction distance (which leads to a retrograde satellite approximately 3.3 times smaller than the prograde one) is not sufficient to assure longer survival times, likely due to the enhanced mass loss induced by disk and bulge tidal shocks.  

\subsubsection{Debris kinematics}\label{sec:debkin}
As commented in Sec.~\ref{sec:res_sa}, the slope of the radial velocity curve nearby the Anti-center $d v_{\rm rad}/d l , l=180^\circ$ can be well reproduced either by low eccentricity, prograde orbits with peri-galacticon at $l\simeq 180^\circ$ or by high eccentricity, retrograde orbits with apo-galacticon at $l\simeq 180^\circ$. Hence, the information provided by the radial velocity curve is not sufficient to determine the rotational sense of motion of the tidal stream progenitor. We must remark on the notably low observational velocity dispersion, which is better reproduced by the prograde model, as one would expect from semi-analytical results (see middle panel of Fig.~\ref{fig:prosa}).

Proper motions in the latitudinal direction do not break the degeneracy between the two different orbital senses of motion, either, since both models predict a similar range of $\mu_b$ values, consistent with the observations taking into account the large error bars. 
In contrast, proper motion constraints in the Galactic plane direction, $\mu_l$, settle the sense of rotation to be prograde ( Fig.~\ref{fig:proper}, lower panel). The prograde model predicts proper motions in that region of the order of $\mu_l\sim -5,0 $ mas/yr, whereas for the retrograde model we find a much larger azimuthal angular velocity $\mu_l\sim 5,20 $ mas/yr due to the larger relative velocity between the Sun and the dwarf galaxy. 
In order to conclusively differenciate the orbital sense of motion, we use proper motions of 10 stars in the stream derived by Munn et al. (2004). Comparing these measurements with the theoretical predictions we find that, except for two stars, observed values of $\mu_l$ are remarkably well reproduced by the prograde model (in magnitude as well as in sign).

\begin{figure}
\plotone{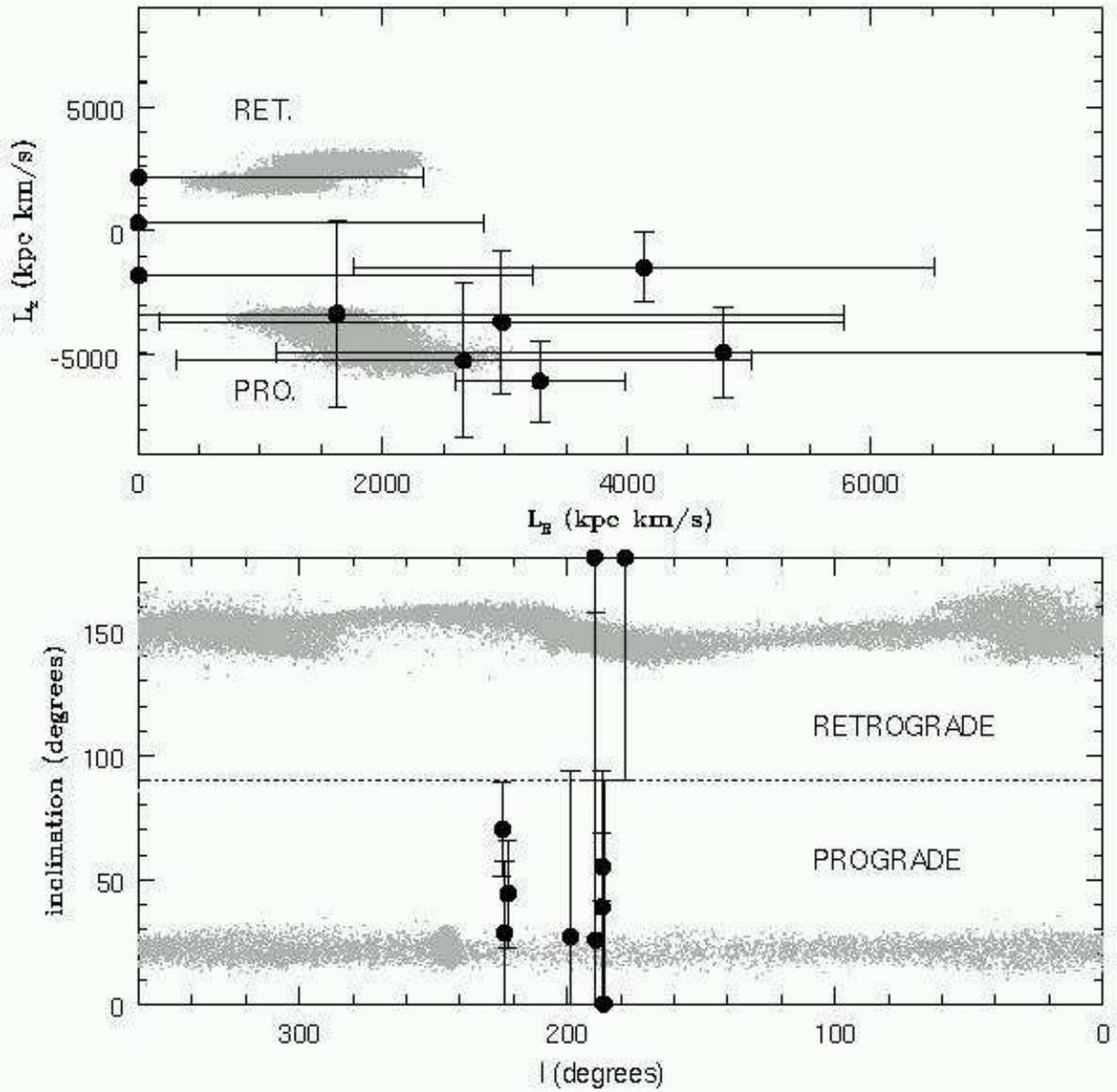}
\caption{ Upper panel: Prograde ({\em pro1} model) vs. retrograde ({\em ret1} model) values of the angular momentum components. Dots represent observational values after being corrected (see text). Lower panel: Orbital inclination as a function of Galactocentric longitude. }
\label{fig:angmom}
\end{figure}

We have transformed $v_r,\mu_l,\mu_b$ into Cartesian velocities in the Galaxy frame (eq.~\ref{eqn:vel}) in order to determine the components of angular momentum and the orbital inclination $\cos i=-L_z/L$, where $L^2=L_R^2+L_z^2$ (bottom panel
\footnote{We note that the observational values of $L_R$ are positive definite quantities with large errors, which leads to a biased estimate unless some statistical correction is applied. We have corrected them by using the technique of Wardle \& Kronenberg (1974).}). 
In Fig.~\ref{fig:angmom} we plot the distribution of N-body particles (models {\em pro1} and {\em ret1}) and observational points in the angular momentum plane $L_R-L_z$ (upper panel). Tidal stream particles are located in well-defined regions determined by the main orbital inclination ($L_z<0$, $L_z>0$ for prograde and retrograde orbits, respectively) and by the main eccentricity (which decreases for decreasing values of $L_R$). In the bottom panel we plot the orbital inclination as a function of Galactocentric longitude ($i<90^\circ$ prograde orbits, $i>90^\circ$ retrograde orbits) which shows that the Monoceros stream progenitor likely follows a prograde orbit. Unfortunately, the observational errors are too large to provide an estimation of the main orbital inclination from these stars.\\
It is interesting to note that simply by measuring kinematical properties of stream stars (radial velocities plus proper motions) we would be able to determine, not only the sense of motion of the progenitor system, but also its orbital eccentricity and inclination if those measurements were accurate enough.  

\subsubsection{Geometrical distribution of debris}\label{sec:debgeo}
As Fig.~\ref{fig:angmom} shows, the kinematical properties of tidal stream particles are similar to those of the progenitor during several periods. As a result, the spatial distribution of debris is an approximate representation of the dwarf galaxy orbit. In Fig.~\ref{fig:proper} we show the projection in the Galactic X-Y plane of dwarf galaxy N-body particles for models {\em pro1} and {\em ret1}, observing that:\\
 (i) Stripped particles from the prograde satellite are located preferentially at two Galactocentric distances, $r\simeq 12 $kpc and $r\simeq 22 $kpc, building two nearly concentric ``rings'' in space. That peculiar spacial distribution forms due to 1) the low orbital eccentricity and 2) the anisotropic mass loss. As Piatek \& Pryor (1995) found, disrupting dwarf galaxies become elongated due to the action of tidal fields. Stripped particles escape preferentially parallel to the major axis which, at the same time, is oriented perpendicular to the density gradient vector. Due to this process, particles belonging to the trailing and the leading tail move, respectively, outwards and inwards with respect to the main body orbit. This process can be clearly seen in top-left panel, the satellite centre-of-density being located at $(X,Y)=(-15,-15)$ kpc and moving clockwise.

The retrograde orbit forms a rosette with apo-galacticon at 22 kpc and peri-galacticon at 7.3 kpc, also providing a reasonably good fit to the Monoceros star detections but failing to reproduce the gap between the two tails observed in the Tri/And regions.

\subsection{Position of the main body}\label{sec:mb}
There are some facts that point to the possible survival of the Monoceros stream's progenitor. Stars escaping from a tidally disrupting system follow the main system orbit for some time (the exact evolution of their orbits depend on several parameters), building up what has been defined as ``tidal tails''. Stars that were lost several orbital periods ago follow orbits in the host galaxy potential nearly independent of their parent system. These particles would spread in a large volume of space (no over-density signature) with no velocity gradient. Helmi et al. (2003) find that only careful measurements of the phase-space structure might indicate whether they belonged originally to a disrupted body. According to their predictions, the spatial location of debris forming a typical tidal stream structure as well as their well-defined radial velocity curve indicate that the tidal tails that we observe are possibly {\em young} and, therefore, either (i) the main body has not been yet completely destroyed or (ii) the disruption occurred recently.

As shown in Section~\ref{sec:res_sa}, the range of distances and Galactocentric latitudes where parts of the stream are detected provides robust values of the orbital inclination and eccentricity of the progenitor. However, the exact vertical and radial distribution of debris depends on the location of the main body (or, equivalently, the time when we stop the N-body simulation).

We have two main constraints to fix the progenitor location: (i) geometrical distribution of debris and (ii) different metalicities observed in different regions. 

\begin{figure*}
\plotone{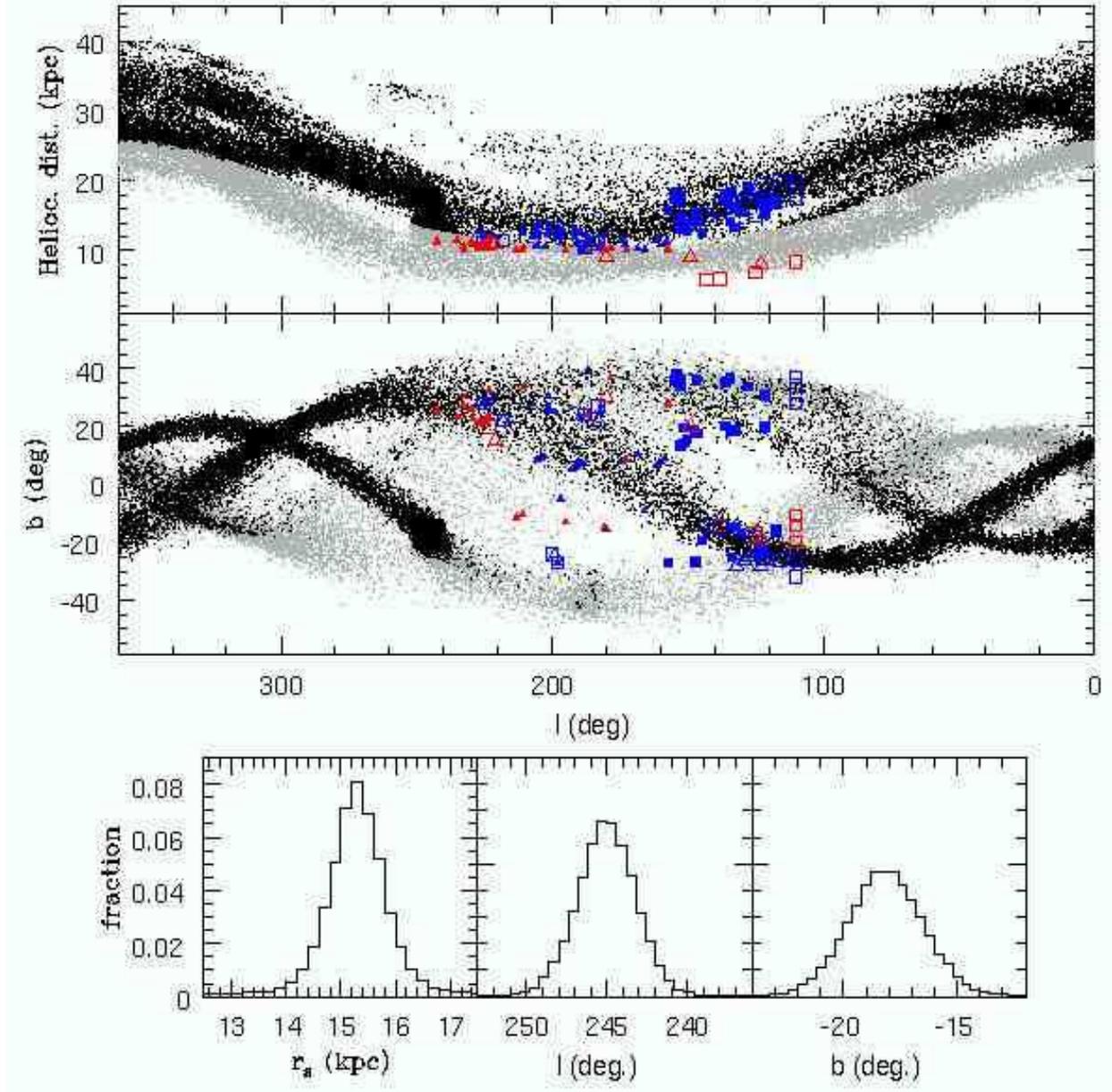}
\caption{ Model {\em pro1} integrated 2.99 Gyr. Heliocentric distance (upper panel) and galactocentric latitude (middle panel) of debris as a function of galactic longitude.  We use notation of Fig.~\ref{fig:proper} to distinguish different observational sources. Grey and black dots denote, respectively, particles at $r\geq 18$ kpc (``outer ring'') and  $r< 18$ kpc (``inner ring'') from the Galaxy centre. Lower panels show the distribution of debris around the main system remnants. From left to right we plot the heliocentric distance, Galactoncentric longitude and the Galactocentric latitude distributions, which are normalized to the total number of dwarf galaxy stars ($N_s=10^5$).}
\label{fig:pro}
\end{figure*}

\vskip0.3cm
 (i) {\bf Geometrical constraint}:\\
 Detections obtained so far show that the Monoceros stream forms a complex vertical structure from the disk plane, if all detections belong to a single tidal stream. As shown in the bottom panel of Fig.~\ref{fig:pro}, such structure can plausibly arise from a single disrupted satellite. 
Different studies (Majewski et al. 2004, Newberg et al. 2004) have reported the presence of two parts of the stream at $l\simeq 110^\circ$, both located at similar projected position but at different heliocentric distances. 
Observations close to the disk plane do not show presence of the distant tail ($b\in [-20^\circ,20^\circ]$), likely due to dust abortion, up to $b>20 ^\circ$ where it is again detected. We remark that the close tail has been detected only at negative latitudes ($b\in [-20^\circ,-10^\circ]$).\\
We mainly use observational points at $110^\circ \leq l \leq 130^\circ$ to determine the progenitor position, since the range of distances of detections at $l\in [140^\circ,240^\circ]$ is considerably smaller which, therefore, provides a weaker constraint. Looking at this longitudinal range we observe that, if  the main body is placed at $l>100^\circ$ (integration time $t<2.66$ Gyr) we find no distant tail at $-25^\circ\leq b \leq -20^\circ$ (only leading tail particles moving in the closest ``ring'' can be found in this region), whereas if placing the main body at $l<200^\circ$  (integration time $t>3.04$ Gyr) we find a luminous close tail at $b>20^\circ$, not present in observations. 
Therefore, we obtain comparable fits to observations if the main body is located in the range $100^\circ\geq l \geq 200^\circ$, which corresponds to integration times from 2.66 Gyr to 3.04 Gyr, respectively (note that prograde orbits move with $dl/dt<0$).

\begin{figure}
\plotone{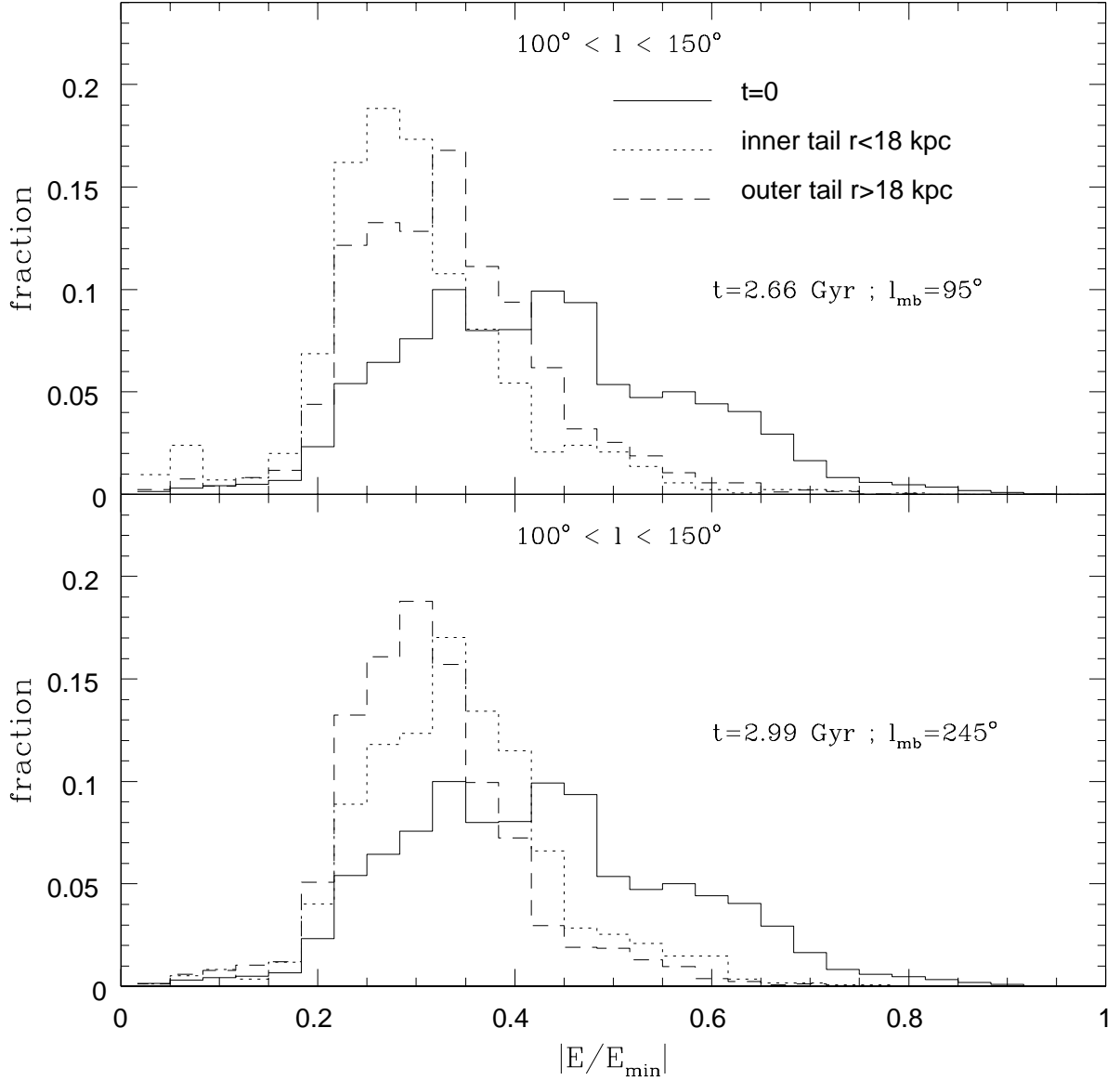}
\caption{ Distributions of initial binding energies. We compare the initial distribution (solid lines) with that at t=2.66 Gyr (upper panel) and t=2.99 Gyr (lower panel) for particles in the range $100^\circ \leq l \leq 150^\circ$. For each integration time we distinguish between particles in the ``inner ring'' ($r<18$ kpc from the Galaxy centre, dotted lines) and  in the ``outer ring''($r>18$ kpc, dashed lines). At t=2.66 Gyr and t=2.99 Gyr the main body is located at $l_{\rm mb}=95^\circ$ and  $l_{\rm mb}=245^\circ$, respectively. We assume that particles with initial high binding energy $|E/E_{\rm min}|\sim 1$ would present higher metalicities than those with low binding energy at $t=0$, $|E/E_{\rm min}|\sim 0$.}
\label{fig:metal}
\end{figure}
\vskip0.3cm
(ii) {\bf Metallicity constraint}:\\
Detections of the distant tail at $100^\circ \leq l \leq 150^\circ$ show metalicities considerably lower than those of the close tail in the same longitudinal range (Rocha-Pinto et al. 2003).
Although we have no information on the initial metallicity gradient of the progenitor, we assume that: 1) stars with originally high binding energy are those with the highest metallicity, as star formation
nd metallicity enrichment happen more intensely in the center of dwarf galaxies (Pagel \& Edmunds 1981, Harbeck et al. 2001)  and 2) those processes generating or enhancing any metallicity gradient have stopped, or are very slow, in the last 3 Gyr\footnote{This assumption is made in order to compare consistently the initial and the final distribution of binding energies. New episodes of star formation during the tidal disruption of the dwarf galaxy would likely occur in the central regions of the system (thus, increasing the metallicity gradient), which does not alter the conclusions of our comparison}. \\
Under those assumptions, the actual distribution of stream metalicities is related to the age of the tidal tails, since low metallicity stars with low binding energy are stripped out by the action of tidal forces more rapidly than stars with initially high binding energy and high metallicity, which will be preferentialy located in recently stripped tails.\\
 Matching the metallicity observed in different regions of the stream reduces the range of possible progenitor locations obtained from geometrical and kinematical constraints. We have compared relative metalicities between the distant and the close tails within $100^\circ \leq l \leq 150^\circ$. In Fig.~\ref{fig:metal} we plot the distribution of stream particles as a function of their initial binding energy for two different positions of the main body (equivalently, two final integration times). Solid lines show the initial distribution (once the satellite is numerically relaxed), dotted and dashed lines the distribution of particles with $r<18$ kpc (``inner ring'') and $r>18$ kpc (``outer ring''), respectively. \\
The upper panel shows the result of placing the satellite at $l=95^\circ$. We can see that the number of particles with low initial binding energy (which one would expect to be metal-poor) is larger in the close tail, contradicting observations. If we integrate a longer time, so that the main body locates at $l=245^\circ$, the number of particles in the distant tail with initial low binding energies is clearly larger than in the close tail. In this case, the outer tail has in average lower metallicity than the close one.

It is interesting to note that the ideal mass loss process, in which mass shells are progressively removed, is fairly approximate. Shell crossing occurs during the evolution of the satellite, likely due to the action of Galactic tidal forces and shocks. As a result, we observe the presence of particles with initially high binding energy in both stream tails.  This fact might explain observations of high and low metallicity stars in same fields (e.g Rocha-Pinto et al. 2003). Although some mixing occurs, Fig.~\ref{fig:metal} shows that the number of particles with initial high binding energy that remain in the main system is larger than in the tidal streams.

In Fig.~\ref{fig:pro} we plot the heliocentric distance (upper panel) and Galactocentric latitude (middle panel) of debris as a function of Galactocentric longitude. Lower panels show the distribution of particles around the main system position. The combination of geometrical and metallicity constraints fixes the main body at $l\sim 245\pm 3$ deg., $b\sim -18\pm 2$ deg., corresponding to an integration time of $t=2.99$ Gyr. The heliocentric distance is approximately $15.2\pm 0.8$ kpc.

We remark that this result is not definitive. Unfortunately, the available data are not yet sufficient to provide strong constrains on the main system location. The progenitor position presented here is fairly approximate and is based on the assumption  of an initial metallicity gradient of the progenitor.
We expect that future detections, mapping larger areas of the sky, will provide stronger geometrical constraints and reduce the number of possible scenarios.

\section{Discussion}\label{sec:dis}
In this Section we discuss the possible identification of the Monoceros stream progenitor in Canis Major (Martin et al. 2004a), by comparing its orbital properties to those of our model {\em pro1}.
We also analyze the orbits of three globular clusters with measured proper motions thought to be associated to CMa dwarf.
Lastly, we briefly summarise and discuss alternative explanations for the Monoceros tidal stream that can be found in the literature.

\subsection{The Triangulus/Andromeda stream}\label{sec:perand}
\begin{figure*}
\plotone{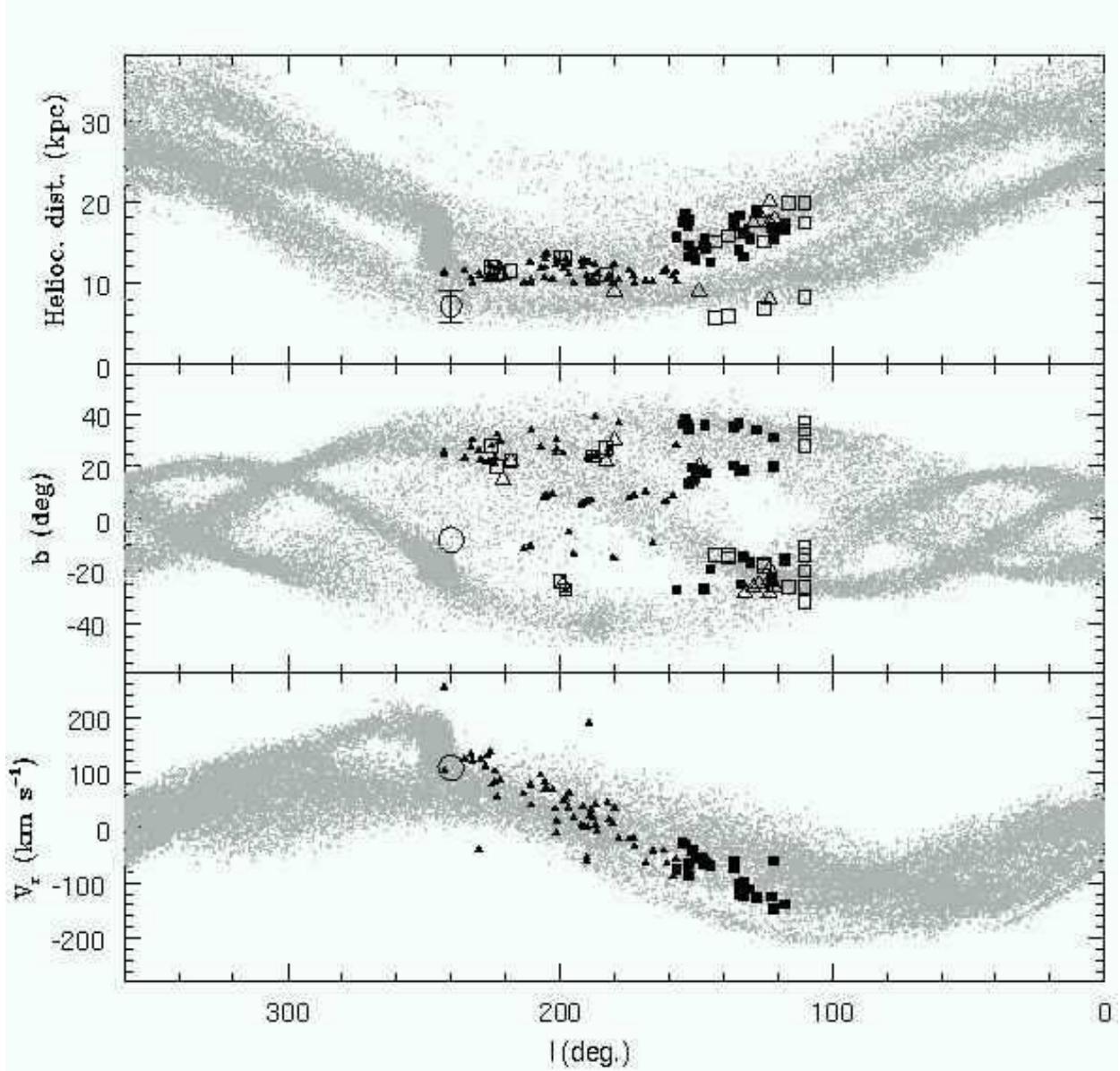}
\caption{Comparison of our model for the Monoceros stream progenitor against CMa dwarf properties (open circle). Upper panel: heliocentric distances of stream particles from model {\em pro1} (grey) against observations. Middle panel: Projection in Galactocentric coordinates. Lower panel: heliocentric radial velocities.}
\label{fig:pertri}
\end{figure*}

In Fig.~\ref{fig:pertri} we show the location of the recent detected Tri/And tidal streams (full squares, Rocha-Pinto et al. 2004) against the debris distribution of model {\em pro1} (grey). We also show those detections used to constrain the orbital parameters. The location and radial velocity of CMa over-density region is shown by a large open circle.
The prograde model suggests the recently discovered stream in Triangulus/Andromeda as natural part of the Monoceros stream, both fitting accurately to the modeled kinematics and spatial distribution of debris. The observed Tri/And streams ($l\in[120^\circ,150^\circ]$) appear as a connection between the Monoceros stream ($l\in[160^\circ,240^\circ]$) and the new SDSS and Ibata 
detections at $l\in[110^\circ,140^\circ]$, justifying the inclusion of these data in our fit, as we discuss in Section~\ref{sec:obs}.

\subsection{The Canis Major dwarf as the progenitor of the Monoceros tidal stream}\label{sec:cma}
Martin et al. (2004a) identified an elliptical over-density of M-giant stars in the Canis Major region with properties suggestive of a disrupted dwarf galaxy: (i) standard models of the Milky way cannot account for the large number of red giants in that region of our Galaxy and (ii) the low dispersion of the radial velocity distribution (approximately 20-25 km/s) is unexpected for a disk population. From the number of M-giant stars, these authors estimate a mass of $10^8-10^9 M_\odot$, similar to the Sgr dwarf galaxy. The location in Galactocentric coordinates is $220^\circ \leq l \leq 260^\circ$,$-15^\circ \leq b \leq -7^\circ$, with no data for $b>-7^\circ$ due to dust abortion. 
In a second paper, martine et al. (2004b) find that the heliocentric distance of this systems is $r_s \simeq 7.2\pm 2$ kpc, with the maximum surface density being located at $l=240^\circ$ and $b=-8.8^\circ$. They also measure a the radial velocity of $v_{\rm r}=109$ km/s.  Adopting the selection criteria of Bellazzini et al. (2004) and Martin et al. (2004a) to select M-giant stars from the CMa field, Momany et al. (2004) measure the following proper motions $(\mu_l,\mu_b)=(-3.5 \pm 2, -0.1 \pm 2)$ mas/yr.

Martin et al. (2004a) concluded that CMa is a satellite galaxy undergoing tidal disruption and, due to its apparent similar position, they suggested that it probably is the remnant of the Monoceros stream progenitor. Here we compare the orbital properties of CMa to our best fit of the Monoceros stream progenitor in order to analyze a possible common origin of both systems. We must remark that our comparison is still fairly preliminar since it goes beyond the original scope of this paper.

\begin{figure}
\plotone{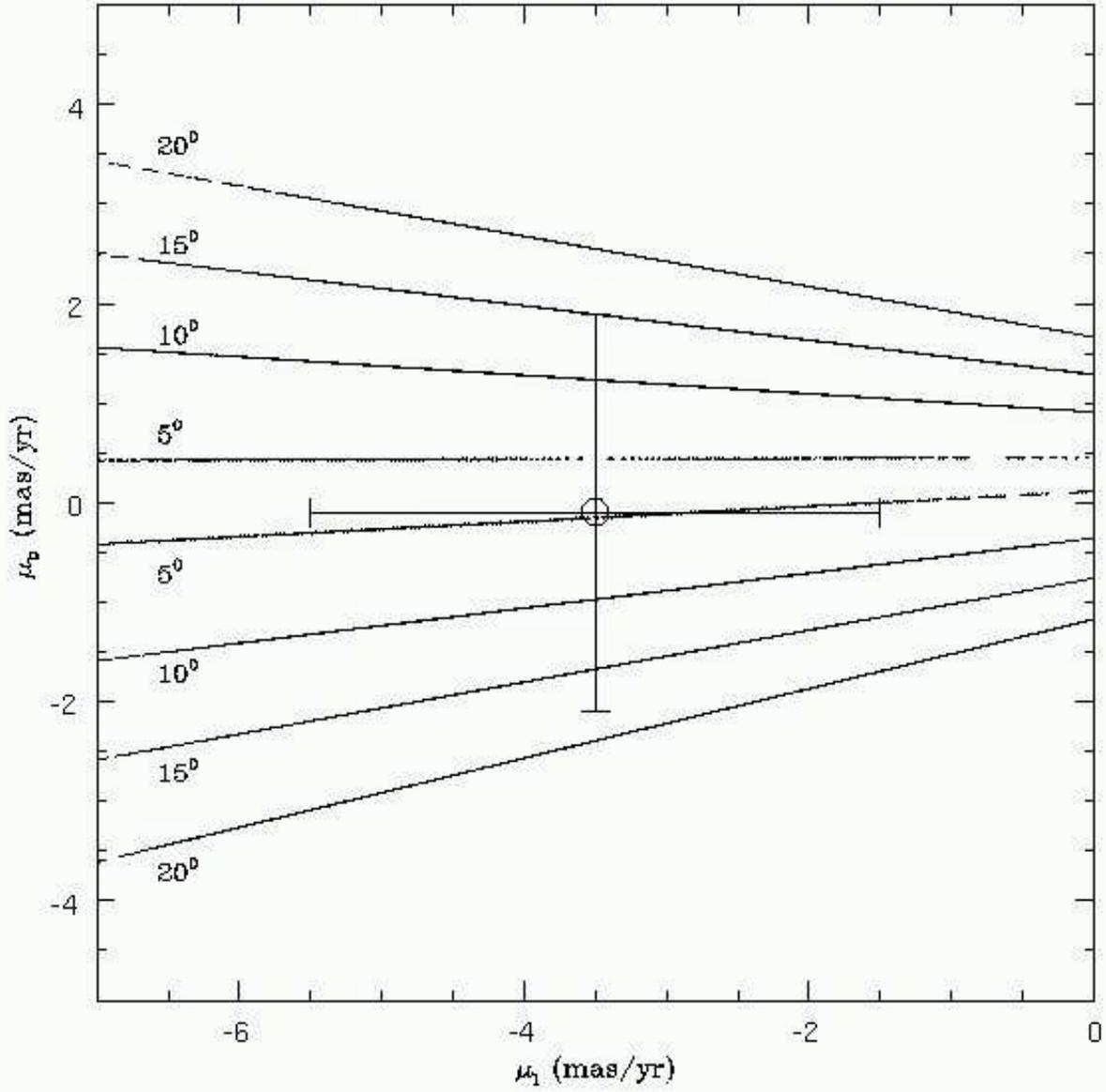}
\caption{Orbital inclination of CMa as a function of $\mu_l,\mu_b$, fixing the radial velocity to 109 km/s and the heliocentric distance to $r_s=7.2$ kpc.}
\label{fig:imu}
\end{figure}

\subsubsection{Orbit calculation} 
We have integrated the orbit of CMa back in time in order to compare its kinematical properties with those of our model. The velocity of CMa in Cartesian coordinates with origin in the Galactic center was obtained using the following expression:
\begin{eqnarray}
\dot{\bf r}=\dot{\bf r}_\odot + v_{\rm rad}(\cos b \cos l,\cos b \sin l, \sin b) + \\\nonumber
 r_s\mu_l \kappa(-\cos b\sin l,\cos b\cos l,0) + \\\nonumber
r_s\mu_b \kappa(-\sin b \cos l,-\sin b \sin l, \cos b)
\label{eqn:vel}
\end{eqnarray}
where $\dot {\bf r}_\odot=(10.0, 225.2, 7.2)$ km/s (Binney \& Merrifield 1998) is the solar velocity, $r_s$ the heliocentric distance and $\kappa\simeq 4.74$ a conversion factor from (kpc mas/yr) to (km/s). The velocity vector is (-145.4,193.2,-4.2) km/s. The resulting CMa's orbit has low inclination ($i=4^\circ$) and is nearly circular $e\simeq 0.16$, as argued by Momany et al. (2004). 
Fig.~\ref{fig:imu} shows that the value of $i$ is strongly sensitive to the errors in proper motions. Taking into account the errors estimated by these authors we obtain that, within $1\sigma$ error, the orbital inclination of CMa lies in the range $i=[0^\circ, 18^\circ]$ if fixing the radial velocity to $v_{\rm r}=109$ km/s and the heliocentric distance to 7.2 kpc. Comparing this result with the best-fitting orbits that we present in this work one can see that:\\
(i) The orbital eccentricity is fairly similar to the one we find (model {\em pro1} predicts $e=0.10\pm 0.05$).\\
(ii) The orbital inclination of CMa is too low to account for the large vertical dispersion of stream stars. This result is, however, not conclusive since a similar orbit to that of model {\em pro1} lies within $1.5\sigma$ error. More thorough measurements of proper motions are necessary to clarify the orbit of the CMa dwarf.\\
(iii) The geometrical and metallicity constrains that we can impose with the available observational data fix the final position of our best-fitting model at $l\simeq 245^\circ$, $b\simeq -15^\circ$ and a distance to the Sun of $r_s\simeq 15$ kpc. CMa appears in a similar direction as the main system of model {\em pro1} but 7 kpc closer. If CMa proves to be the Monoceros stream progenitor, a way to reconcile the close distance of the CMa with the distant stream detections might be found by increasing the progenitor mass. A simple estimation of the decay rate induced by dynamical friction on nearly circular orbits is  $\Delta r \propto M_s r \Delta t$ (eq. 7-25 of Binney \& Tremaine 1986). For the model {\em pro1} $\Delta r\simeq 3$ kpc after $\Delta t=3$ Gyr. Imposing $\Delta r\simeq 9$ kpc in 3 Gyr leads to an initial mass of the progenitor of approximately $M_s'/M_s=\Delta r'/\Delta r=9/3=3$, i.e $M_s'\simeq 1.8 \times 10^9 M_\odot$. However, it is unclear whether the tidal debris of such a massive satellite would also reproduce the observed distribution. In order to check this assumption, new investigations adopting the CMa dwarf as the Monoceros stream progenitor appear necessary, which goes beyond the scope of this paper.

\begin{figure}
\plotone{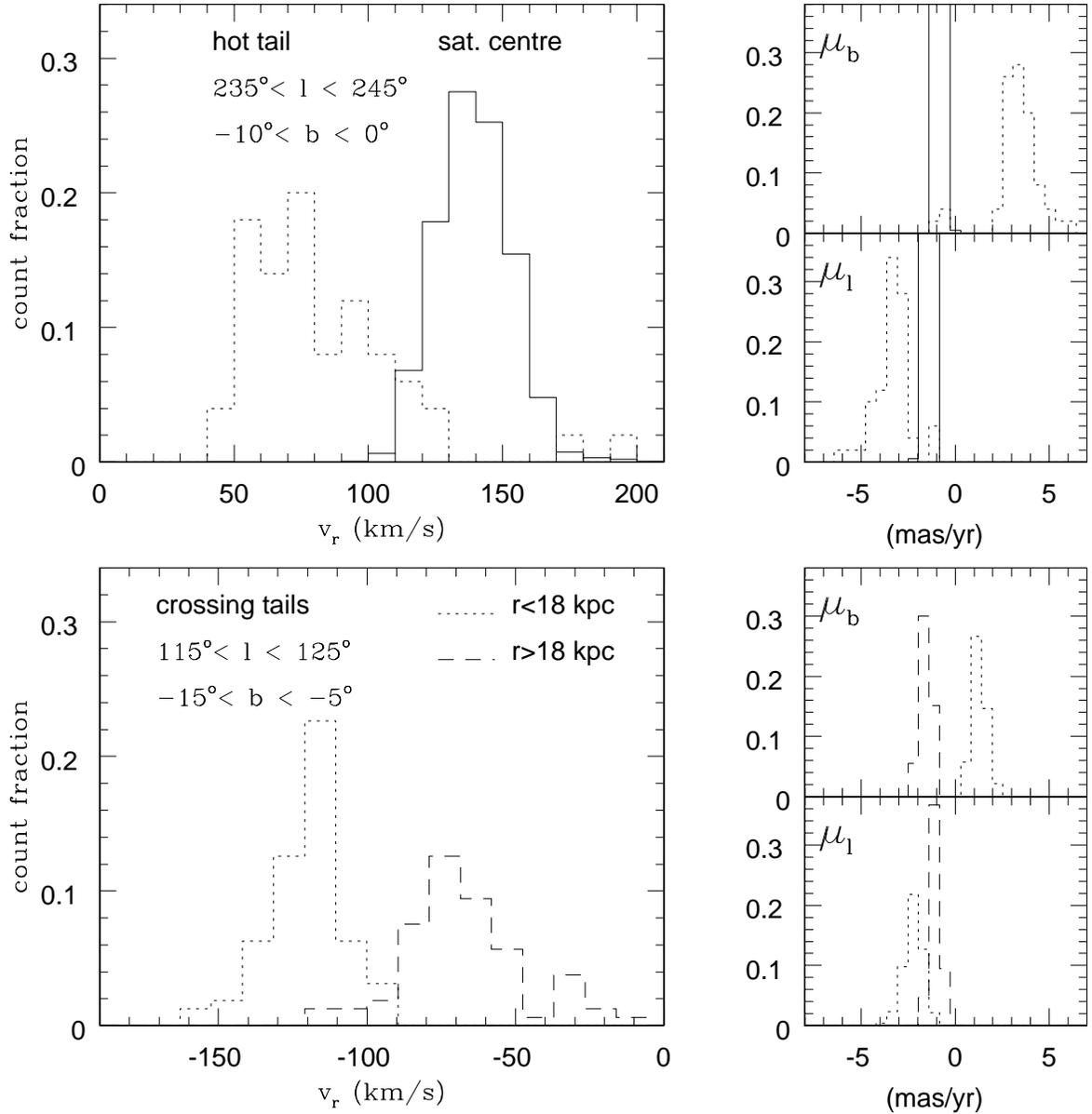}
\caption{ Left column: Distribution of radial velocities in the CMa region from model {\em pro1} (upper panel) and in a region of tail orverlap (lower panel). The velocity distribution of the satellite center includes particles within $243^\circ\leq l \leq 247^\circ$ and $-17 \leq b \leq -13^\circ$. The number of particles in each distribution was 8400 in the dwarf center and 50 in the surrounding tail.
Dotted and dashed lines in the lower panel show the distributions from particles in the inner and outer ``rings'', respectively. Right column: Distribution of proper motions in the regions indicated above.}
\label{fig:vr}
\end{figure}

\subsubsection{3D velocity distribution in the CMa central region.}
We have also analyzed the distribution of radial velocities and proper motions in different regions of the stream in order to contrast them to the recently available observational values.\\
As we can observe in Fig.~\ref{fig:pertri}, our model predicts a hot tail overlapping the progenitor remnants. This tail was stripped out approximately 0.6 Gyr ago and belong to the trailing tail (i.e to the ``inner ring''). In upper panel of Fig.~\ref{fig:vr} we plot the radial velocity distribution (dotted line) of 50 particles located in the neighbourhood of the main system's projected position and compare it to that of the center of our satellite model (full line, note that each distribution is separately normalized to the number of particles in the sample). This Figure shows than the velocity distribution of both tails can be clearly differentiated, with peaks at $v_r\simeq 135$ km/s (cold tail) and $v_r\simeq 75$  km/s (hot tail). 
Although some particles of the underlying tail are included in the velocity distribution of the satellite centre, one cannot appreciate a second peak because the surface density of the hot tail is much lower that of the satellite centre. In lower panel of this Figure, we repeat the calculations in a crossing tail region, where the surface densities of both tails are comparable. In this case, one observes clearly two distributions of radial velocities (those from the distant and the close tail) in the same particle sample.\\
The observed low-dispersion bimodality cannot be reproduced by standard models of the Milky Way and is likely a proof of a stream detection, as we show here.

Intriguingly,  Martin et al. (2004b) report a bimodal velocity distribution in CMa stars located at the center of the CMa dwarf. Measuring the radial velocity of 27 M-giant stars in one degree radius around CMa position they observe two peaks in the radial velocity at $v_r\simeq 63$ km/s (10 stars) and $v_r\simeq 109$ km/s (17 stars) with very low dispersion (around 5 km/s and 11 km/s, respectively). 
These values are around 20\% larger than those of model {\em pro1}.
Extrapolating the resulting distribution of our model to their detections suggests that in the CMa region two stream tails are overlapping. Moreover, the fact Martin et al. (2004b) measure similar number of stars in both peaks indicates that their surface density must be similar (if we assume that the number of M-giants in a given stream is proportional to its surface density). This, however, contradicts the presence of a dwarf galaxy in Canis Major since the surface density at the center of these objects is several orders of magnitude larger than in the tidal streams. 

That paradox might be solved if their small sample of M-giant stars was strongly contaminated with stars that belong to the hot tail surrounding CMa. In that case, the proper motions provided by these authors might be also affected. In right column of Fig.~\ref{fig:vr} we show the distribution of proper motions in the regions indicated above. As in the distribution of radial velocities, different stream parts lead to well differentiated curves. Looking at the region where the dwarf remnants are located, we can observe that the main system shows a fairly narrow distribution, with maxima at $(\langle \mu_l\rangle,\langle \mu_b\rangle) \simeq (-1.7,-0.6)$ mas/yr, whereas the distribution of the hot, overlapping stream tail is centred at $(\langle \mu_l\rangle,\langle \mu_b\rangle) \simeq (-3.4,3.4)$ mas/yr. Thereferore, including hot tail stars in the dwarf remanant sample (in a significant proportion) would lead to a smaller $\langle\mu_l\rangle$ and a larger $\langle\mu_b\rangle$, which results in a lower orbital inclination (see Fig.~\ref{fig:imu}).

\subsection{Possible associated stellar clusters}\label{sec:cluster}
Frinchaboy et al. (2004) have collected a set (15) of globular and open cluster that show a trend in their radial velocity curve as well as in spatial location which may indicate that those systems were stripped from a satellite in disruption process.  

Martin et al. (2004a) claimed that four of those clusters (NGC 1851, 1904, 2298 and 2808) belong to a globular cluster system associated with the CMa dwarf, arguing for the possible detection of the Monoceros stream's progenitor in Canis Major.

We have earlier shown that radial velocities and positions do not provide sufficient information to distinguish between pro- and retrograde orbital motions. In order to break that degeneracy, accurate measurements of proper motions are needed.
In this Section we test the possible association of those clusters with measured proper motions (NGC 1851, 1904 and 2298 from Dinescu, Girard \& van Altena et al. 1999) with the CMa dwarf.
Additionally, we also discuss the possible common origin of six ``possibly associated clusters'' suggested by Frinchaboy et al. (2004) by contrasting locations and radial velocities with the predictions of our best-fitting model. Unfortunately, no proper motions are available for those systems, so that the results shall not be conclusive.

\begin{figure*}
\plotone{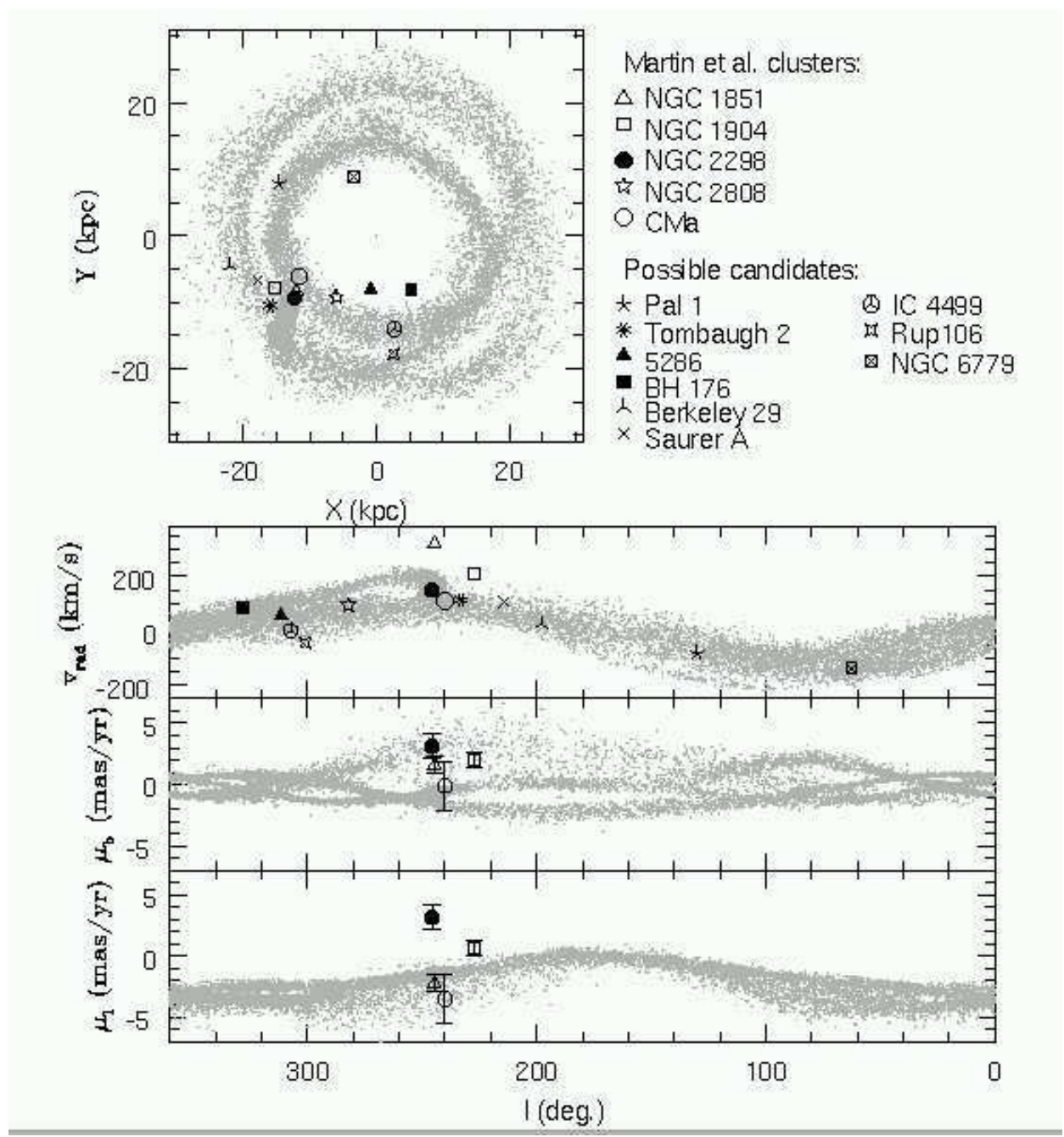}
\caption{ Upper panel: X-Y projection of the {\em pro1} model debris against some globular and open clusters possibly associated with the Monoceros stream. Middle-upper panel: radial velocity curve. Lower-middle panel: Proper motions in the latitudinal component. Lower panel: Proper motions in the longitudinal component. The open circle denotes the position and radial velocity of the CMa dwarf (Martin et al. 2004b). }
\label{fig:cum}
\end{figure*}

In Fig.~\ref{fig:cum} we plot the X-Y projection of model {\em pro1} (upper-right panel) and clusters listed in the upper-left panel. Middle-upper panel shows the radial velocity, middle-lower and lower panels plot proper motions in the longitudinal ($\mu_l$) and latitudinal ($\mu_b$) directions, respectively. This Figure shows that:
\vskip0.3cm
(i) {\bf CMa clusters}:\\
None of these globular clusters present kinematical properties consistent with those of the CMa dwarf. \\
Firstly, NGC 1851 and NGC 1904 have radial velocities of approximately 320 km/s and 208 km/s whereas that of the CMa dwarf is approximately  109 km/s (Martin et al. 2004b). After integrating the orbit we find that both globular clusters move on orbits with much higher eccentricities, $e=0.7$ and $e=0.65$, respectively (in agreement with Dinescu, Girard \& van Altena et al. 1999, who use a a similar Galaxy potential). On the other hand, the orbital eccentricity of escaping particles remains fairly similar to that of the main body (see Fig.~\ref{fig:angmom}), leading to a well-defined radial velocity curve. If one assumes that this holds for stripped globular clusters, we come to the conclusion that NGC 1851 and 1904 are unlikely associated with CMa. Additionaly, the orbital inclinations of those systems are considerably higher than that of CMa.\\ 
The third globular cluster with measured proper motions, NGC 2298, follows a retrograde, highly eccentric orbit ($e\simeq 0.78$, Dinescu, Girard \& van Altena et al. 1999), arguing against any association with CMa and the other clusters.
\\
Unfortunately, there are no proper motions available for NGC 2208 and, therefore, insufficient information to allow a determination of its possible association with CMa.

\vskip0.3cm

(ii) {\bf Other clusters}:\\
In upper-middle panel we can see that these clusters present a trend in radial velocities shown by Frinchaboy et al. (2004) consistent with our model. Their projected positions (upper panel) also appear to form a ``stream'' in space. Searching through the Galactic globular cluster sample provides some additional, plausable candidates to be associated with the Monoceros stream: Rup 106, IC 4499 and NGC 6779, from a comparison of their projected position, distance and radial velocity with the predictions of our model.   
Unfortunately, whether or not these clusters are associated to the Monoceros stream cannot be asserted in absence of proper motions.

\subsection{Thick disk stars or tidal stream debris?}\label{sec:disk}
The main selection criteria used to discriminate between Milky Way stars and stars of an ``external'' origin were: (i) stream stars are observed to be in overdense regions not predicted by our standard Milky Way model and (ii) the turn-off stars appear blue, old and metal-poor, characteristics similar to those of the Sagittarius dwarf galaxy stars (Newberg et al. 2002). Those conditions by themselves do not exclude other explanations for the observations.\\
Here would like to comment briefly on two possible origins of the observed tidal streams: (i) thick disk and (iii) a satellite disruption.

In Section~\ref{sec:res} we have shown that the stream detections can be reproduced by a disrupting satellite on a prograde, low-inclination, nearly circular orbit. The kinematical as well as spatial distribution of debris resemble what one would expect for a thick disk stars (see Fig.~\ref{fig:proper}) except for one point: it is difficult to reconcile the large vertical dispersion ($\Delta z= 2\sin i \times 24.5$ kpc$\simeq 20 $kpc) and the large distances ($\sim 20 $kpc from the Galaxy centre) of proposed stream stars with a disc-like distribution.  To clarify this point, we have carried out a simple experiment: we have calculated the probability of finding thick disk stars at the Monoceros stream location. The number of stars between $m$ and $m+dm$, where $m$ is apparent magnitude can be written as
\begin{equation}
N(m)d m= N (r_s) \frac{dm}{dr_s} dr_s= N(r_s)\frac{5}{\ln 10}\frac{d r_s}{r_s},
\label{eqn:Nm}
\end{equation}
where $r_s$ is the distance of a star to the Sun. We have used the relationship $m-M=5\log_{10}(r_s)$ to convert distances into apparent magnitudes. The probability of finding a star between $r_s$ and $r_s+d r_s$, in the range of solid angles $\Omega,\Omega+d\Omega$ is
\begin{equation}
N(r_s,\Omega)d r_s d \Omega= \rho(r_s,\Omega) r_s^2 d r_s d\Omega,
\label{eqn:Nl}
\end{equation}
where $\rho(r_s,\Omega)$ is the thick disk density distribution in the solar frame. Combining both equations one has that the probability function is
\begin{equation}
P(r_s,\Omega)= A\rho(r_s,\Omega) r_s ,
\label{eqn:P}
\end{equation}
where $A$ is some normalization constant.
The fraction of thick disk stars in a given solid angle is therefore
\begin{equation}
N(r_s,\Omega)= \frac{\int_0^{r_s} \rho(x,\Omega) x dx}{\int_0^\infty \rho(x,\Omega) x dx} .
\label{eqn:N}
\end{equation}

\begin{figure}
\plotone{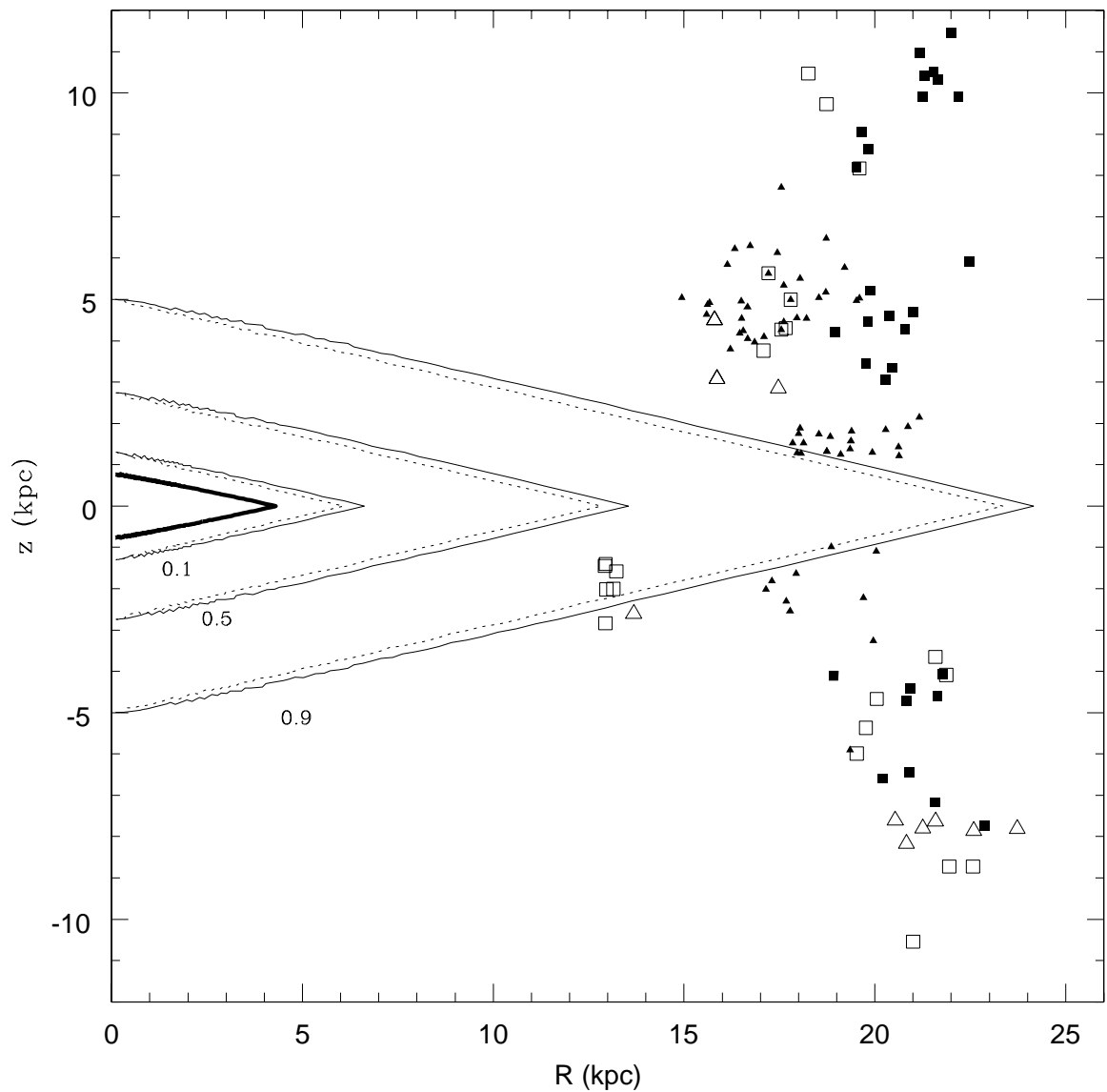}
\caption{Iso-contours of the thick disk number of particles against stream detections. The Sun is located at $R=z=0$. The strong line represents the maximum like-hood directions of finding thick disk stars towards the Galactic anticenter ($l=180^\circ$).  Lines show the $R,z$ values where the number of thick disk stars is 10\%, 50\% and 90\% of the total number if looking at the Anti-center (full lines) and $70^\circ$ away (dotted lines).}
\label{fig:prob}
\end{figure}

As we can see, $P$ is zero for $r_s=0$ and $r_s\rightarrow \infty$ so that, for a given solid angle, there is a distance where the probability finds a maximum.
In Fig.~\ref{fig:prob} we plot the maximum like-hood iso-contour of observing thick disk stars (strong full line) in the Galactic anticenter direction (eq.~\ref{eqn:P}). The Sun is placed at $R=z=0$. Full lines show the iso-contours of the thick disk fraction of particles as a function of their position in cylindrical coordinates when looking in the Galactic anticenter direction (eq.~\ref{eqn:N}).  Dotted lines show the iso-contours if looking $70^\circ$ away (i.e, $l=110^\circ$ or $l=250^\circ$) in order to take into account the range of Galactic longitudes where the stream has been observed. In both cases we plot the 10\%, 50\% and 90\% contours.
As we can see, the Galactic anticenter is the direction for which the number of distant thick disk stars is maximum. For this plot, we have used the thick disk model proposed by Chen et al. (2001), which follows exponential profiles in the planar and vertical directions with scale-lengths of 3.5 kpc and 0.75 kpc, respectively. \\
 This Figure shows that stream detections presented in this work are unlikely related to the thick disk population because (i) they are located in a narrow range of co-planar distances ($R$), as one would expect for debris from a disrupting galaxy in a nearly circular orbit; (ii) they present a large vertical dispersion, of approximately 20 kpc which cannot be reconciled with a disc-like structure and (iii)  they have been detected in positions where the fraction of thick disk stars is $\leq$10\% of the total number of stars in that direction. The only doubtful data might be found in the colour-magnitude detections at $R\simeq 13$ kpc, which lie within 90\% fraction of thick disk stars. These points correspond to what our model identifies as the ``close ring'', which presents a higher metallicity than more distant detections.  \\
It is interesting to note that stars stripped from the Monoceros stream progenitor have not yet contributed to the thick disk population. The decay rate of the dwarf galaxy appears slower (around 3 kpc in 3 Gyr) than the mass loss rate (50\% in the same time period), so that the dwarf galaxy will be likely detstroyed before reaching the inner regions of the thick disc. The mass loss process, however, depends on the initial structural parameters of the dwarf, which were fixed {\em ad hoc} in this work. More accurate measurements of the stream surface brightness will provide better estimations of its survival time.

\section{Conclusions}\label{sec:conc}
In this paper, we have used a combined semi-analytic/N-body technique to explore the nature and origin of the Monoceros tidal stream. This method has allowed us to explore a large parameter space in a systematic way, with the goal of constraining the orbital properties of the Monoceros stream. \\
We have found that the available observational data at the present day is sufficient to robustly constrain some aspects of the Monoceros stream history and progenitor. In particular:  (i) the heliocentric distance range at which the stream is observed plus the heliocentric radial velocity curve determines an  orbital eccentricity of $e\simeq 0.10\pm 0.05$, (ii) the range of galactic latitudes indicates an orbital inclination of $i\simeq 25^\circ\pm 5^\circ$ from the disk plane and, lastly, (iii) proper motions of stream stars are only compatible with a prograde orbit.

We found that it is considerably less straightforward to predict the main body position through numerical calculations. The range of distances, the radial velocity curve and the vertical extension of debris are relatively insensitive to the final location of the satellite galaxy, since they define a given volume in phase-space that the progenitor fills up after several wraps. Yet, the projected positions of stream tails are time-dependent reflecting, therefore, the position of the main system or, equivalently, the final integration time of our N-body model.\\
 Besides the uncertainty in the satellite remnant position, the lack of detections at low galactic latitudes (owing to disk absorption within $b\in [-20^\circ,20^\circ$]) as well as for a complete range of Galactic longitudes makes difficult to constrain the location of the main system. Owing to these limitations, our model matches the geometrical and kinematical distribution of debris if the main system remnants locate within the range $100^\circ\geq l \geq 200^\circ$, corresponding to integration times from 2.66 to 3.04 Gyr. \\
 On the other hand, observations show that close detections of the stream have higher metalicities than the distant ones, which provides an additional constraint to our model.  We have used this constraint by assuming that the progenitor's stellar metalicities are related to their initial binding energy. In particular, we assume that stars with low binding energies move, in average, in the outer satellite regions and, therefore, should present lower metalicities than those with high binding energies. By comparing the initial binding energy (metallicity) distribution with observations for the range of locations commented above, we find that the metallicity gradient between the distant and the close detections can only be reproduced for integration times longer than 2.9 Gyr. In that case, the main system location ($l\sim 245^\circ$, $b\sim -18^\circ$) is similar to that of the Canis Major dwarf, with a heliocentric distance of $15.2\pm 0.8$ kpc.
The treatment employed to determine the distribution of metalicities in the stream is, however, approximate, because have no way of knowing the initial metallicity gradient in the stream progenitor. Moreover, we implicitly assume that detections with different metalicities reflect an initial property of the progenitor. Other scenarios are possible and, for example , we cannot discard that streams with different metalicities at different distances belong to different progenitors.

The best-fitting model predicts a halo axis-ratio of $q_h=0.6$. However, the poorly-constrained selection function of observations and the incomplete area coverage leads to degenerated solutions for different values of $q_h$. In particular, we find that similar fits can be obtained with halo axis-ratios $q_h\in[0.6,0.8]$.

The model we present here is, therefore, far from being definitive. Future detections, mapping larger areas of the sky, will provide more constrains on the progenitor's orbit and a better determination of parameters listed in Table~\ref{tab:param}. New detections are also necessary to constrain the progenitor location by means of numerical models.

With the recently available kinematical data of the Canis Major dwarf, which Martin et al. (2004a) claim to be the progenitor of the Monoceros stream, we have integrated the orbit backwards in time, comparing its orbital properties against those of our model. We find contradictory points in favour and against that suggestion:

(i) In favour:
\begin{itemize}
\item The orbital eccentricity of CMa ($e \simeq 0.16$) and orbital sense of motion (prograde) are consistent with those of our model.
\item The CMa's projected location ($l=240^\circ$, $b=-8^\circ$) is consistent with the geometrical and metallicity constrains imposed by the Monoceros stream stars. 
\item The orbital inclination ($i\simeq 4^{+14}_{-4}$ deg.) is consistent with that of our model within $1.5\sigma$ errors from proper motions.
\item The radial velocity of CMa ($v_r=109$ km/s) is also similar to that of our model ($v_r=135$ km/s) and presents a low dispersion incompatible with thick disk features.
\end{itemize}

(ii) Against:
\begin{itemize}
\item The orbits of globular clusters NGC 1851, 1904 and 2298, which Bellazzini, Ferraro \& Ibata (2003) and Martin et al. (2004a) claim to be an evidence that the CMa over-density region is the remnant of the Monoceros stream progenitor, are inconsistent with the orbit of CMa, since NGC 1851 and NGC 1904  move on highly eccentric orbits and NGC 2298 on a retrograde, eccentric one.

\item The bimodality in the velocity distribution of CMa's M-giant stars (Martin et al. 2004b) can be reproduced by our model in regions where the projected positions of two stream tails overlap. That bimodality in $v_r$ cannot be observed in the central part of our satellite model since the surface density is several orders of magnitude larger than that of overlapping stream tails. This fact seems to point out to a possible contamination of background giant stars in the sample used to measure proper motions and the radial velocity of the main system. As a result, the orbital inclination might have been underestimated.
\item The main body of our best-fit
model's progenitor is $\sim 15$ kpc from the Sun --- twice as distant as
the observed Canis Majoris stellar structure. Since the mass of our model is poorly constrained, larger mass values cannot be rejected. In particular, dynamical friction would drive our satellite model down to 7.2 kpc in 3 Gyr for an initial mass of approximately $1.8 \times 10^9 M_\odot$, whereas our best-fitting model has an approximate initial mass of $(6\pm 3)\times 10^8 M_\odot$. It remains unclear, however, whether the resulting distribution of debris from such a massive satellite undergoing tidal disruption would also reproduce the observations.
\end{itemize}

 Unfortunately, with the results obtained in this work we cannot unambigously determine whether the orbit of the CMa dwarf is consistent with that of the Monoceros stream progenitor, partly due to the uncertainty in the available observational data. 
More strict criteria to select stars that belong to the CMa dwarf appear necessary to obtain reliable measurements of the kinematical properties of this system.

\vskip1cm

We want to thank Dana I. Dinescu for helping us to analyse the information obtained from proper motions. We also thank H. Rocha-Pinto and J. D. Crane for giving us access to their data before being published. 

Funding for the creation and distribution of the SDSS Archive has been
provided by the Alfred P. Sloan Foundation, the Participating
Institutions, the National Aeronautics and Space Administration, the
National Science Foundation, the U.S. Department of Energy, the
Japanese Monbukagakusho, and the Max Planck Society.  The SDSS Web
site is (\texttt{http://www.sdss.org/}).

The SDSS is managed by the Astrophysical Research Consortium (ARC) for the
Participating Institutions. The Participating Institutions are The University
of Chicago, Fermilab, the Institute for Advanced Study, the Japan Participation
Group, The Johns Hopkins University, Los Alamos National Laboratory, the
Max-Planck-Institute for Astronomy (MPIA), the Max-Planck-Institute for
Astrophysics (MPA), New Mexico State University, University of Pittsburgh,
Princeton University, the United States Naval Observatory, and
the University of Washington.


{}

\end{document}